%% file: main.tex
\documentclass[9pt,twocolumn,twoside]{osajnl}

\journal{ao} 

\newcommand{\mrm}[1]{\mathrm{#1}} 

\usepackage{xcolor}
\usepackage{tablefootnote}
\usepackage{siunitx}
\DeclareSIUnit{\beli}{Bi}
\DeclareSIUnit{\dBi}{\deci\beli}

\usepackage{subfigure}

\setboolean{shortarticle}{false}

\title{The Simons Observatory: Modeling Optical Systematics in the Large Aperture Telescope}
\input{Author.tex}
\affil[*]{Corresponding author: jon@fysik.su.se}

\begin{abstract}
We present geometrical and physical optics simulation results for the Simons Observatory Large Aperture Telescope. This work was developed as part of the general design process for the telescope; allowing us to evaluate the impact of various design choices on performance metrics and potential systematic effects. The primary goal of the simulations was to evaluate the final design of the reflectors and the cold optics which are now being built. We describe non-sequential ray tracing used to inform the design of the cold optics, including absorbers internal to each optics tube. We discuss ray tracing simulations of the telescope structure that allow us to determine geometries that minimize detector loading and mitigate spurious near-field effects that have not been resolved by the internal baffling. We also describe physical optics simulations, performed over a range of frequencies and field locations, that produce estimates of monochromatic far field beam patterns which in turn are used to gauge general optical performance. Finally, we describe simulations that shed light on beam sidelobes from panel gap diffraction. 
\end{abstract}

\begin{document}

\maketitle

\input{1Introduction/introduction.tex}
\input{2OpticalDesign/opticaldesign.tex}

\input{3CameraRayTrace/cameraraytrace.tex}
\input{4Spillover/spillover.tex}
\input{5PhysicalOptics/po.tex}

\input{6PanelGaps/6panelgaps.tex}

\input{Conclusion/conclusion.tex}

\section{Funding Information}
This work was supported in part by a grant from the Simons Foundation (Award \#457687, B.K.). JEG acknowledges support from the Swedish National Space Agency (SNSA/Rymdstyrelsen) and the Swedish Research Council (Reg. no. 2019-03959). GEC acknowledges support from NSF GRFP. MD thanks the Simons Observatory and The University of Pennsylvania. RD thanks CONICYT for grant BASAL CATA AFB-170002. GF acknowledges support from the European Research Council under the European Union’s Seventh Framework Programme (FP/2007-2013) / ERC Grant Agreement No. [616170], and support by the UK STFC grants ST/T000473/1. FM acknowledges the support from the World Premier International Research Center Initiative (WPI), MEXT, Japan. PDM contribution is supported by the Simons Observatory and funds from the Arizona State University Interplanetary Initiative. KM acknowledges support from the National Research Foundation of South Africa. MDN acknowledges support from NSF award AST-1454881. This manuscript has been authored by Fermi Research Alliance, LLC under Contract No. DE-AC02-07CH11359 with the U.S. Department of Energy, Office of Science, Office of High Energy Physics.

\section{Disclosures}
The authors declare no conflicts of interest.

\bibliography{references}

\end{document}

%% file: Author.tex
\author[1]{Jon E. Gudmundsson}
\author[2]{Patricio A. Gallardo}
\author[3]{Roberto Puddu}
\author[4]{Simon R. Dicker}
\author[1]{Alexandre E. Adler}
\author[5]{Aamir M. Ali}
\author[6]{Andrew Bazarko}
\author[7]{Grace E. Chesmore}
\author[4]{Gabriele Coppi}
\author[8]{Nicholas F. Cothard}
\author[1]{Nadia Dachlythra}
\author[4]{Mark Devlin}
\author[3]{Rolando Dünner}
\author[9]{Giulio Fabbian}
\author[10]{Nicholas Galitzki}
\author[11]{Joseph E. Golec}
\author[6]{Shuay-Pwu Patty Ho}
\author[12]{Peter C. Hargrave}
\author[4]{Anna M. Kofman}
\author[5, 12]{Adrian T. Lee}
\author[4]{Michele Limon}
\author[14]{Frederick T. Matsuda}
\author[15]{Philip D. Mauskopf}
\author[16, 17]{Kavilan Moodley}
\author[18]{Federico Nati}
\author[2, 19]{Michael D. Niemack}
\author[4]{John Orlowski-Scherer}
\author[6]{Lyman A. Page}
\author[20]{Bruce Partridge}
\author[21]{Giuseppe Puglisi}
\author[22]{Christian L. Reichardt}
\author[11]{Carlos E. Sierra}
\author[23]{Sara M. Simon}
\author[10]{Grant P. Teply}
\author[12]{Carole Tucker}
\author[24]{Edward J. Wollack}
\author[4]{Zhilei Xu}
\author[4]{Ningfeng Zhu}

\affil[1]{The Oskar Klein Centre, Department of Physics, Stockholm University, SE-106 91 Stockholm, Sweden}
\affil[2]{Department of Physics, Cornell University, Ithaca, NY 14853, USA}
\affil[3]{Instituto de Astrof\'isica and Centro de Astro-Ingenier\'ia, Facultad de F\'isica, Pontificia Universidad Cat\'olica de Chile, Macul, Santiago, Chile}
\affil[4]{Department of Physics and Astronomy, University of Pennsylvania, Philadelphia, PA 19104, USA}
\affil[5]{Department of Physics, University of California, Berkeley, CA 94720, USA}
\affil[6]{Joseph Henry Laboratories of Physics, Jadwin Hall, Princeton University, Princeton, NJ 08544, USA}
\affil[7]{Department of Astronomy and Astrophysics, The University of Chicago, Chicago, IL 60637, USA}
\affil[8]{Department of Applied and Engineering Physics, Cornell University, Ithaca, NY 14853, USA}
\affil[9]{Department of Physics \& Astronomy, University of Sussex, Brighton BN1 9QH, UK}
\affil[10]{Department of Physics, University of California San Diego, La Jolla, CA 92093, USA}
\affil[11]{Department of Physics, University of Chicago, Chicago, IL 60637, USA}
\affil[12]{School of Physics and Astronomy, Cardiff University, Cardiff, CF24 3AA, UK}
\affil[13]{Physics Division, Lawrence Berkeley National Lab, USA}
\affil[14]{Kavli Institute for the Physics and Mathematics of the Universe (WPI), The University of Tokyo Institutes for Advanced Study, The University of Tokyo, Kashiwa, Chiba 277-8583, Japan}
\affil[15]{School of Earth and Space Exploration and Department of Physics, Arizona State University, Tempe, AZ 85281, USA}
\affil[16]{Astrophysics Research Centre, University of KwaZulu-Natal, Westville Campus, Durban 4041, South Africa}
\affil[17]{School of Mathematics, Statistics \& Computer Science, University of KwaZulu-Natal, Westville Campus, Durban 4041, South Africa}
\affil[18]{Department of Physics, University of Milano-Bicocca, Milano, Italy}
\affil[19]{Department of  Astronomy, Cornell  University, Ithaca, NY 14853, USA}
\affil[20]{Department of Astronomy, Haverford College, Haverford PA 19041, USA}
\affil[21]{Computational Cosmology Center, Lawrence Berkeley National Laboratory, Berkeley, CA 94720, USA}
\affil[22]{School of Physics, University of Melbourne, Parkville, VIC 3010, Australia}

\affil[23]{Fermi National Accelerator Laboratory, MS209, Batavia, IL 60510, USA}
\affil[24]{Goddard Space Flight Center, Greenbelt, MD 20771, USA}

%% file: 1Introduction/introduction.tex
\section{Introduction}


The Simons Observatory is a next-generation cosmic microwave background (CMB) experiment that will be deployed in the Atacama Desert at an altitude of 5200 meters \cite{Galitzki2018}. The experiment will be composed of one large aperture telescope (LAT) and three small aperture telescopes (SATs). Although the experiment will cover a wide range of science goals within mm-wavelength astrophysics \cite{SO2019}, two of the primary science goals include: 1) improving limits on the amplitude of a hypothesized primordial gravitational wave background produced in the early universe; and 2) mapping the matter distribution of the universe by measuring the integrated gravitational deflection of the CMB. In order to reach its science goals, the experiment needs multi-frequency coverage with polarization sensitivity. The Simons Observatory will have 6 frequency bands with band centers spanning \mbox{27--\SI{270}{\giga\hertz}} with detectors divided into 3 dichroic bands termed LF, MF, and UHF (see Table~\ref{tab:telescope_deployment}). The number of detectors in each band is set to achieve the required noise levels needed for these science goals.

As CMB polarization experiments become increasingly sensitive, the relative importance of systematic effects grows larger. A significant class of potential systematics are caused by non-idealities in the optical systems for CMB measurements. In this paper, we describe a suite of optical simulations used to inform the design choices for the Simons Observatory Large Aperture Telescope. These analyses have helped identify and remedy potential sources of systematic error and detector loading. The current results from this work, presented in this paper, shed light on the ability of this experiment to reach its science goals. 


Section \ref{sec:design} describes the general optical design. Section \ref{sec:go} contains basic ray tracing analysis used to optimize the design as well as analysis used to inform the cryogenic baffling for optics tubes. Section \ref{sec:spillover} discusses ray tracing analysis used to mitigate warm spillover past both secondary and primary mirrors. Section \ref{sec:po} summarizes basic results from physical optics analysis, including estimates for far field beam size and ellipticities and cold optics spillover. Section \ref{sec:panels} describes simulations used to predict beam sidelobes from panel gap diffraction.

%% file: 2OpticalDesign/opticaldesign.tex
\begin{figure*}[t!]
    \centering
    \includegraphics[width=0.8\linewidth]{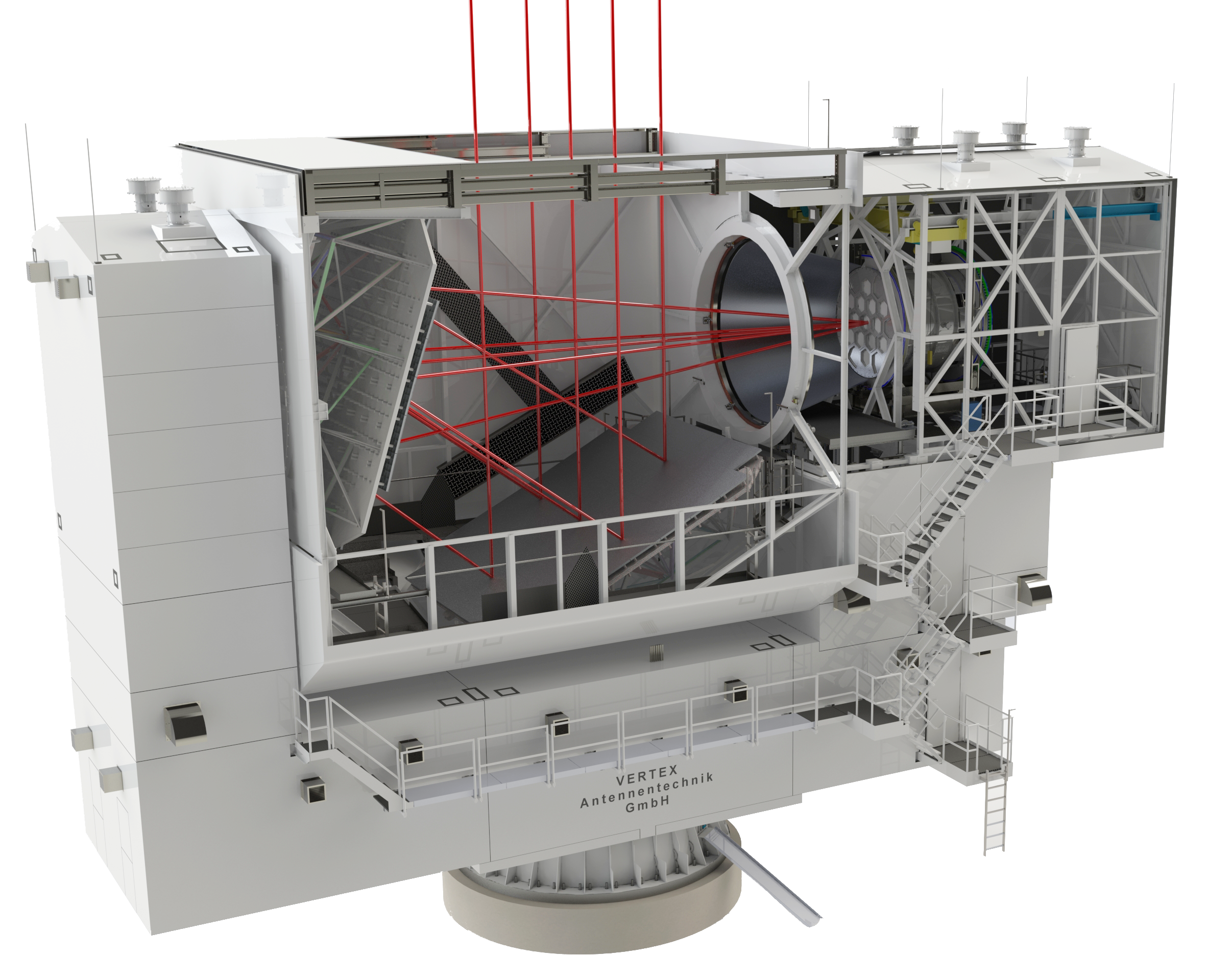}
    \caption{Rendering of the LAT, courtesy of Vertex, with some of the telescope cladding removed to show the placement of the mirrors and the cryogenic receiver. The elevation structure, shown here pointing towards the zenith, houses the primary and secondary mirrors and is designed to rotate in elevation from \ang{-90} to \ang{0} on the other side (\ang{270} throw). The cylindrical cryogenic receiver can rotate to follow the rotation of the elevation structure. In a time-reverse sense, rays from the receiver hit the secondary and then the primary before going through the aperture on the elevation housing and out to the sky (e.g., see red ray bundle coming from the center optics tube). The projected diameter of the primary is \SI{6}{\meter}. A conical forebaffle (discussed in Section \ref{sec:spillover}) is mounted in front of the receiver. Part of the baffle is suppressed in this rendering so as to not block the front of the cryogenic receiver.}
    \label{fig:LAT_3D}
\end{figure*}    

\begin{figure}[]
    \centering
    \includegraphics[width=0.5\linewidth]{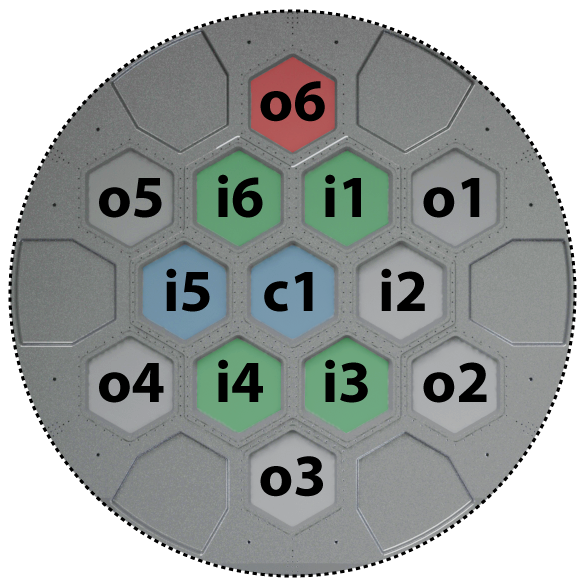}
    \caption{Cropped version of a LAT rendering showing the LAT cryogenic receiver as viewed from the secondary mirror. The center optics tube is labeled as \textsf{c1}, while tubes in the inner (\textsf{i}) and outer (\textsf{o}) circles are labeled as \textsf{i1}, \textsf{o1}, \textsf{i2}, \textsf{o2}, etc. The cryostat is \SI{2.4}{\meter} in diameter. Initial deployment will only populate 7 of the 13 optics tubes, with: two UHF in \textsf{c1} and \textsf{i5} (blue hex); four MF in \textsf{i1, i3, i4, i6} (green hex); and one LF in \textsf{o6} (red hex).}
    \label{fig:LAT_label}
\end{figure}    


\section{Optical \& Mechanical design}
\label{sec:design}
\subsection{The telescope optics}
The Simons Observatory's Large Aperture Telescope is a Crossed-Dragone telescope with an approximately \SI[number-unit-product=\text{-}]{6}{\meter} mean entrance pupil diameter and a mean effective focal length of \SI{15.6}{\meter}. It shares a common optical and mechanical design with the CCAT-prime telescope and is currently under construction by Vertex Antennentechnik in Germany before transportation and installation at the Simons Observatory site in Chile.\footnote{Vertex Antennentechnik GmbH, \url{https://www.vertexant.com/}} The Crossed-Dragone design \cite{Grimes2009, Bischoff2013, Hileman2010, Sugai2020, LSPE2020} supports large and unblocked optical throughput while meeting the Mizuguchi-Dragone condition \citep{Mizuguchi1978, Dragone1978} to minimize cross-polar response. Additionally, both the primary and secondary mirrors are perturbed from the basic conic sections used in a classic Crossed Dragone telescope design in such a way as to cancel first order coma in off-axis fields. This allows for a diffraction limited field-of-view (FoV) of the telescope of approximately \ang{7.8} effective diameter at \SI{90}{\giga\hertz}, thus meeting the Simons Observatory's requirement for optical throughput. A rendering of the telescope with a basic ray-trace is shown in Fig.~\ref{fig:LAT_3D}. Figure~\ref{fig:LAT_label} shows a view of the LAT cryogenic receiver as seen from the secondary mirror with labels for each of the 13 optics tubes. Further details of the optical design can be found in \cite{Parshley2018, Parshley2018b}. In general, the optical design is based on work described in \cite{Niemack2016}. Although all 13 tubes can and have been studied with the analysis tools presented in this paper, our discussion will focus on the optical performance of the center tube and the 6 tubes in the inner circle of the cryogenic receiver (see  Fig.~\ref{fig:LAT_label} and Table \ref{tab:telescope_deployment}). We therefore cover the performance of the MF and UHF bands, but do not discuss LF beam performance in this paper. 



The primary and secondary mirrors of the Simons Observatory LAT are composed of 77 and 69 individual rectangular panels (Al 5083), respectively, each covering approximately \SI{0.5}{\meter\squared} and weighing about \SI{7}{\kilogram}. The panels are designed to have a \SI{1.2}{\milli\meter} gap at \SI{-6}{\celsius} with full gap closure at \SI{60}{\celsius}. Areas surrounding the optically active regions on both the secondary and primary can be lined with panels that would reflect receiver sidelobe power to the sky in order to minimize thermal loading on detectors; we currently have no plans to deploy such panels. For the same reason, however, a roughly 3-m long conical baffle with a 9-deg half opening angle is placed in front of the telescope receiver (see Fig.~\ref{fig:LAT_3D} and Section~\ref{sec:spillover}). 

The required system half wavefront error (HWFE) is < \SI{35}{\micro\meter} rms \citep{Parshley2018b}. Contributions to the overall surface HWFE budget for the primary and secondary reflector system have been estimated in several categories, including: a) individual panel manufacturing errors (\SI{10.2}{\micro\meter}); b) manufacturing margin on the alignment of the carbon fiber support structures for the primary and secondary mirrors (\SI{10.5}{\micro\meter}); c) both types of manufacturing errors, a and b, combined with gravitational deformations (above \ang{30} elevation), wind deformations (up to \SI{6}{\meter/\second}), temperature changes and temperature gradients lead to surface error budgets of \SI{15.3}{\micro\meter} and \SI{15.0}{\micro\meter} for the primary and secondary, respectively. In addition to budget items a-c, we estimate contributions to the HWFE budget from: d) relative alignment between the mirrors under environmental conditions similar to c (\SI{5.0}{\micro\meter}); and e) mirror panel alignment errors.  Assuming photogrammetry is used to align the mirror panels, the panel alignment errors are estimated to be \SI{20}{\micro\meter}, which should result in a system HWFE of approximately \SI{30}{\micro\meter}. If better alignment techniques are used, such as the holography and laser metrology systems planned for CCAT-prime \citep{Parshley2018b}, the panel alignment errors could be decreased substantially, potentially decreasing the system HWFE to as low as about \SI{22}{\micro\meter}. Existing measurements of individual panel surface rms together with engineering models suggest that the HWFE requirement will be met.


As shown in Fig.~\ref{fig:LAT_3D}, the two mirrors are mounted in an enclosing elevation structure with the telescope focus outside the rotating housing. The receiver sits on a Nasmyth platform. To reduce systematic effects, such as the change in beam shape with boresight rotation, the receiver is held on a mount designed to co-rotate with the telescope elevation structure. The elevation structure can be pointed from \ang{-90} elevation through to \ang{0} on the other side (\ang{270} throw). The telescope can therefore be pointed straight down in order to protect it from the elements. 



\subsection{Cold optical design}
The cold re-imaging optics that couple detectors to the telescope are shown in Figs.~\ref{fig:LAT_3D} and~\ref{fig:cold_optics}. The design can accommodate 13 optics tube modules, each containing 3 silicon lenses, a filter stack, and, between \mbox{lenses 2 and 3}, a Lyot stop cooled to \SI{1}{\kelvin}. The filter stack consists of thin IR blocking filters at cryogenic temperature stages of \SI{300}{\kelvin}, \SI{80}{\kelvin}, and \SI{40}{\kelvin}. Additionally, there is an absorbing alumina filter at \SI{80}{\kelvin} and four thicker low pass quasi-optical filters with two at \SI{4}{\kelvin}, and one each at \SI{1}{\kelvin}, and \SI{100}{\milli\kelvin} \citep{Ade2006, Tucker2006, Pisano2016, Inoue14}. Each tube contains either an LF, MF, or UHF focal plane and, other than their meta-material anti-reflective coatings \cite{ARcoating}, the lens designs do not change between different optics tubes. Apart from optical elements that are obviously frequency dependent (such as bandpass filters), the only optical element that varies between optics tubes is an alumina filter which is given a wedge shape (up to \ang{1}) in off-center optics tubes in order to make the chief ray of the central field of each tube parallel to the axis of the cryostat. This allows all optics tubes to be telecentric with each other and removes the need for telescope dependent axial shifts. Close packing of the tubes, and hence better use of the focal plane, is achieved with the use of hexagonal vacuum windows; the diameter of these windows is approximately \SI{395}{\milli\meter}. More details on the cold optical design can be found in \cite{Dicker2018} and the cryostat in \cite{Ningfeng2018}. 

\subsection{Optics tube design}
By itself, the cold optical design is relatively simple; a wedged filter, three lenses, and a stop represent the key optical elements in each tube. However, the practical implementation of this design, shown in Fig.~\ref{fig:cold_optics}, is far more complex. The bolometers on the focal plane need to be cooled to \SI{100}{\milli\kelvin} in a low magnetic field environment, while the cryogenic cooling at \SI{100}{\milli\kelvin} is limited to only tens of microwatts per optics tube. This leads to significant challenges in thermal and magnetic shielding. In particular, great care is needed in the design of the thermal filter stack, including aspects of cutoff frequencies and heatsinking \cite{Ningfeng2018}. On top of this, mitigation of stray light is critical --- the SO mapping speed is limited in part by thermal loading from the atmosphere as well as loading due to spillover on warm optics. As an example, a one-percent increase in the amount of optical throughput that spills past the warm reflectors (on to \SI{300}{\kelvin}) will reduce the 150-GHz mapping speed by roughly \SI{20}{\percent} \cite{Hill2018}. A considerable effort in reducing warm spillover is therefore warranted. This is the primary goal of the work presented in Sections~\ref{sec:go}~and~\ref{sec:spillover}. 

    
\begin{table}[htbp]
\begin{center}
\caption{\bf Simons Observatory bands and detectors}
\label{tab:telescope_deployment}
\begin{tabular}{cccc}
\hline
Frequency & Bands & \# Tubes$^{\dagger}$ & \# Detectors/tube \\
\hline
LF & 27/\SI{39}{\giga\hertz} & 1 & 600 \\
MF & 90/\SI{150}{\giga\hertz} & 4 & 5184 \\
UHF & 220/\SI{270}{\giga\hertz} & 2 & 5184 \\
\hline
\end{tabular}\\
\end{center}
$^\dagger$\small{Numbers quoted for initial deployment only.}
\end{table}

%% file: 3CameraRayTrace/cameraraytrace.tex
\begin{figure*}[th!]
    \centering
    \includegraphics[width=0.6\linewidth]{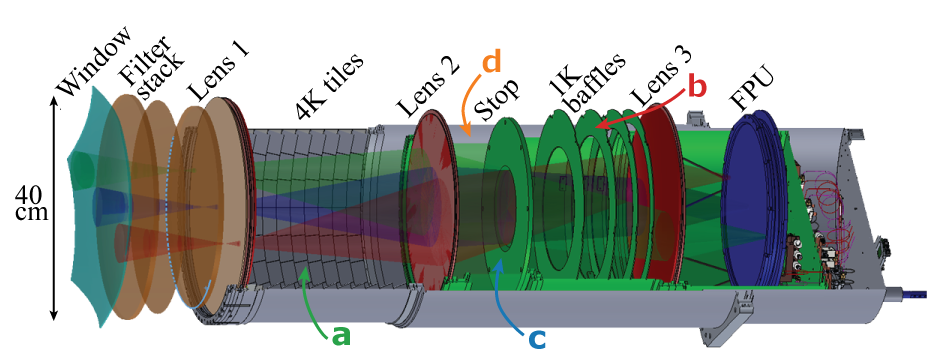}
    \caption{Cross section view through the central optics tube with the different components labeled. Lens 1 is cooled to \SI{4}{\kelvin} while the Lyot stop and Lenses 2 and 3 are cooled to \SI{1}{\kelvin}. The structure includes thermal breaks, magnetic shielding, optical baffling, and mechanical support. Grey structure is \SI{4}{\kelvin}, Green is \SI{1}{\kelvin} and the \SI{100}{\milli\meter} focal plane unit (FPU) is Blue. Lenses/filters are shown in red/brown. Analysis of the impacts of reflections and scattering off the different surfaces is presented in Section~\ref{sec:go}. Some of the areas marked with letters (\textsf{a}-\textsf{d}) are described in more detail in Section~\ref{sec:go}\ref{sec:go2}. The front of the tube (\textsf{a}) is shown with absorbing tiles in place.}
    \label{fig:cold_optics}
\end{figure*}    

\begin{figure}[t!]
    \centering
    \includegraphics[width=0.48\textwidth]{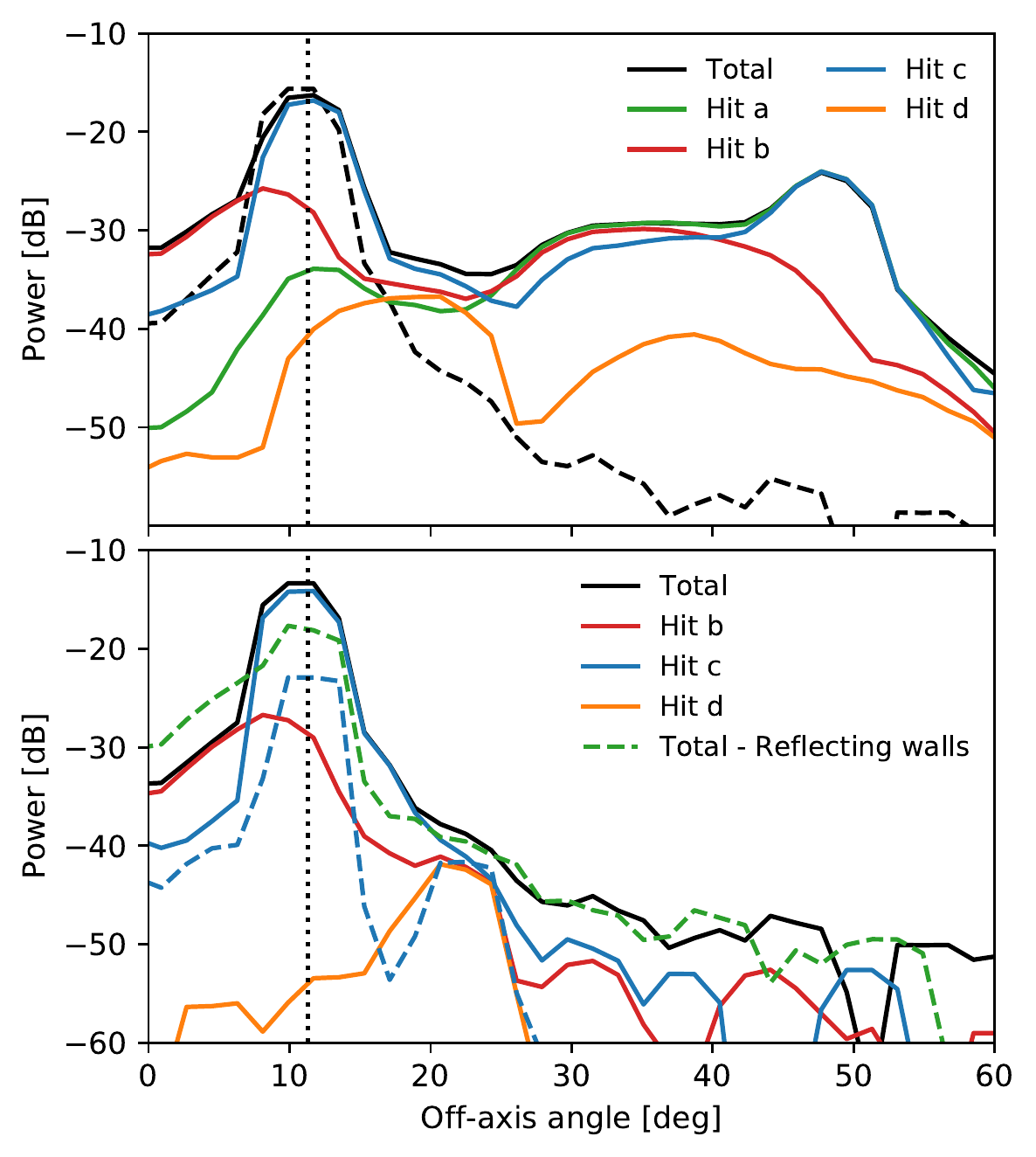}
    \caption{Summary of results from non-sequential ray tracing. Angles that extend beyond the vertical dashed black line will miss the optically active part of the secondary mirror and be far more likely to contribute to loading. Curves are normalized such that the sum of all ray paths traced is normalized to a peak power of 0~dB. The overall power (excluding rays that only interacted with filters and lenses) is shown in black. Other colors show angular power distributions for rays that interact with different parts of the optics tube, marked \textsf{a}-\textsf{d} on Fig.~\ref{fig:cold_optics}. The top panel is for a tube with epoxy/carbon absorber. The dashed black line on this panel is the total if this absorber on the front part of the tube (area a) is replaced by absorbing tiles \citep{Chesmore2020}. The bottom panel shows the contribution of different components with these tiles in place. The dashed blue line shows what happens if you redesign the stop to have a \ang{45} edge while the dashed green line is what happens if you make the back end of the optics tube (area b) \SI{100}{\percent} reflecting --- showing that the baffling of this part of the tube is of less importance. Note that the results shown here are frequency independent.}
    \label{fig:rays_vs_angle}
\end{figure}

\section{Geometric analysis}
\label{sec:go}

\subsection{Model setup}
We used sequential (time-forward) ray tracing in Zemax\footnote{Zemax OpticStudio: \url{https://www.zemax.com/products/opticstudio}} during the original reflector and cold optics design process \citep{Parshley2018, Dicker2018}. While this approach allowed us to optimize the shapes of the lenses and to produce maps of metrics such as the Strehl ratio across the focal planes, this analysis did not account for various optical non-idealities such as stray light (internal reflection) which depends critically on incidence angles. For stray light analysis, we constructed a model of the optics tube including different baffles and anti-reflective coatings using Zemax's non-sequential mode (Fig.~\ref{fig:cold_optics}). Using this approach, we launched randomly generated (time-reverse) rays from the focal plane and calculated the angular distribution of power emerging from the hexagonal cryostat window to determine spillover past the secondary mirror.

For the three lenses, we assumed coatings representing the thickness, number of layers, and refractive index of the as-machined metamaterial \cite{Datta2013, Coughlin2018}. For all filter surfaces, we assumed $\lambda /4$ coatings on all air transitions. For other surfaces, including the structure of the optics tube and the Lyot stop, we applied coatings that were allowed to vary between \SI{100}{\percent} absorbing to \SI{100}{\percent} reflecting. For those surfaces, we studied different types of reflections, including both specular and Lambertian. 
For each combination of surface type and geometry, we had Zemax launch $10^8$ rays from the focal plane at random angles up to \ang{45}. This angle was chosen to be significantly larger than the acceptance angle of the feeds and lenslets on the focal planes and increasing this angle to \ang{65} made no difference to our results. At each surface, the reflected and refracted paths were calculated and each assigned an appropriate fraction of the power from the incident ray (in accordance with surface properties). These paths were traced until they either hit another surface in the model, they missed all surfaces, exited the cryostat window, or the fraction of power they carried dropped below \SI{0.01}{\percent} of the original ray from the focal plane. Increasing this cutoff threshold to \SI{0.03}{\percent} made no statistical difference to the output of the simulation. All rays that made it out of the hexagonal vacuum window were then saved to a database. By summing up the power of the rays in this database then, excluding the effects of diffraction which are discussed in Section~\ref{sec:po}, we can calculate the power exiting the cryostat as a function of angle. This approach also allows us to filter the rays to include those that have (or have not) interacted with different combinations of surfaces, thereby identifying critical areas where the addition of absorbers or changes to geometries are important. 

In this analysis, we concentrate on those ray paths that interacted with something other than lenses and filters. More specifically, we focus on rays that interact with: 1) the region between Lenses 1 and 2 at the front of the optics tube; 2) the cold stop surface; and 3) the \SI[number-unit-product=\text{-}]{1}{\kelvin} baffles between the cold stop and Lens 3. Ray paths that have reflections off any pairs of lens or filter surfaces can produce ghost images that contribute to detector sidelobes. However the only thing that can be done about these is to design the best possible anti-reflective coatings (which is already a design goal).




\subsection{Modeling absorbers}
\label{sec:go2}
We want most cavities inside the optics tubes to be black --- ideally perfectly absorbing for all incidence angles and frequencies. In the past many different approaches have been used (for a review see \cite{ReviewOfCoatings}) to approximate this condition. At millimeter wavelengths a common approach is to paint surfaces with an epoxy mixed with subwavelength conductive particles such as carbon black for absorption, alumina or fuzed silica frit for coeffienct of thermal expansion (CTE) compensation, and the option of adding scattering centers such as wavelength-sized grains of silicon carbide \cite{Diez2000,Klaassen2002}. Such coatings have the advantage that they are inexpensive, easy to apply to complex shapes, and have a CTE suitably matched to the substrates enabling survival during thermal cycling. For these reasons we adopt this type of absorptive coating as the baseline for all metal surfaces.  

Although measurements have shown a very low level of specular reflection ($< \SI{2}{\percent}$) from this coating type, there is a body of evidence that a significant fraction of incident light is scattered \cite{Diez2000,Gallardo2012,Gallardo2018}. This arises from the (0.5 to 10) wavelength scale roughness of the surface -- light which is not absorbed by the coating can be diffusely scattered as well as specularly reflected. For our analysis, a value for this coating of \SI{50}{\percent} absorbing and \SI{50}{\percent} scattering was assumed.  Although pessimistic for normal incidence, rays reflected at oblique angles could easily be fully polarized. The modelling of the wavelength scale roughness of a typical epoxy coating was not possible within Zemax, so for the epoxy coatings the angle of incidence was ignored. However, for all other surfaces (lenses, filters, bare metal walls, and other types of absorbers) smooth dielectric layers were assumed and polarization taken into account when tracing rays.

Other possible blackening materials include microwave absorbers such as HR10 or TK tiles \cite{eccosorb, teraherz}, but these can be harder to apply to curved surfaces such as the inside of the optics tubes and, in the case of HR10, one must contend with fragments of them breaking off. In addition, all absorbers add both weight and cost to the optics tubes. Consequently, the properties of different parts of the optical model were varied between \SI{100}{\percent} reflecting, \SI{100}{\percent} absorbing, and \SI{100}{\percent} scattering. For areas which had little effect on large angle scattered light, we chose to drop the absorbing material (saving cost and weight) while some areas were found to be critical and improvements over the default \SI{50}{\percent} scattering surface were needed. Flat and angled TK tiles were investigated for areas that needed improvement.

\subsection{Results}
Selected results from our simulations are shown in Figure~\ref{fig:rays_vs_angle}. With the initial simulation using the baseline epoxy coatings, we observe scattered light reaching up to \SI{-25}{\decibel} at angles exceeding \ang{20}\ from the axis to the optics tube. At angles greater than $\sim$\ang{50}, an azimuthally symmetric sidelobe of only \SI{-30}{\deci\bel} can have a large enough solid angle that, should all power go to 300K, it can easily add a few Kelvin to the detector loading, degrading sensitivity and introducing significant systematic effects, such as cross-polar sidelobe response. 

Nearly all the rays that contribute to these sidelobes interact with the front of the optics tube (part $a$ in Fig.~\ref{fig:cold_optics}; green line in Fig.~\ref{fig:rays_vs_angle}). Making this area more black is clearly important. A number of possibilities were investigated --- there is little room between the mechanical walls of the tube and the optical path, leaving no room for deep baffles. We found various practical issues associated with traditional absorbers such as HR10. Looking at the database of rays scattered from this surface, we see that most rays hit at oblique angles (peaking at $\sim$\ang{70} from the normal) making most flat absorbers ineffective. We therefore decided on implementing custom angled absorbers.

 When the epoxy absorber in these areas was replaced by angled absorbing tiles as shown in Fig.~\ref{fig:cold_optics} the situation became much better \citep{Chesmore2020}.
Such tiles, which are based on the TK tile concept, can be made relatively cheaply and have on their surface tapered spikes on a sub-wavelength scale to act as a broadband AR coating, reducing any reflections. The Zemax modeling shows that these tiles reduce power scattered to wide angles to less than \SI{-40}{\decibel} at angles greater than \ang{25}.

Even with these tiles in place, there remains significant power at angles less than \ang{25} --- an area which will end up just outside the secondary mirror. Most of this involves a reflection from the edge of the Lyot stop ($c$ in Fig.~\ref{fig:cold_optics}) and then a reflection from one of the many lenses or filters on the sky side of the stop. The original stop was of the classic knife edge design, but close to the center, getting a thick enough layer of absorber without vignetting some of the off-axis fields was challenging. By removing all absorber and making the inner surface reflecting but at \ang{45} from the axis of the tube, this power is instead redirected to the \SI{4}{\kelvin} tube where it is absorbed. Further from the center of the stop, black absorbing TK tiles can be placed.

With the above two changes, simulations predict the dashed blue line at the bottom of Fig.~\ref{fig:rays_vs_angle}. As an experiment, we tested making the walls and baffles ($b$ and $d$ in Fig.~\ref{fig:cold_optics}) \SI{100}{\percent} reflective. Although an increase in the scattered light is clearly visible (the dashed green line), the increase is small. If the baffles are left with epoxy absorber and only the walls of the tube made reflective, very little increase in the scattered light could be seen. This represented a considerable reduction in mass, and so this design has been adopted for the Simons Observatory LAT optics tubes. 

If the assumptions used to generate these simulations are valid, then secondary spillover from the cold optics will be less than \SI{1}{\percent} of the total optical throughput. However, it is quite possible that these simulations are missing important internal reflection mechanisms. Therefore, strategies to mitigate unexpected spillover from the cold optics are discussed next.

%% file: 4Spillover/spillover.tex
\begin{figure}[]
    \centering
    \includegraphics[width=0.4\textwidth]{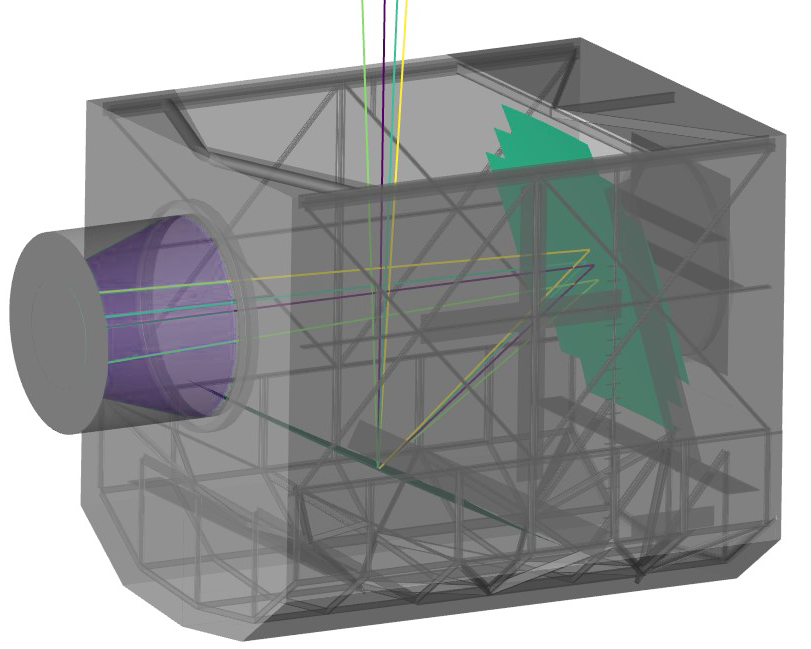}
    \caption{Simplified non-sequential 3D Zemax model used in the spillover analysis. The conical receiver baffle is highlighted in purple. The secondary mirror is shown in green and four principal rays, roughly coinciding with the location of \textsf{c1}, \textsf{i5},  \textsf{i4}, and \textsf{i6}, are also shown.}
    \label{fig:LAT_3D_spill}
\end{figure}    

\section{Spillover and far sidelobes predictions}
\label{sec:spillover}

The cold optics tube design of the ACTpol experiment, which includes three lenses with an intermediate primary image Lyot stop, is similar in many ways to the one adopted for the Simons Observatory \cite{thornton2016atacama}. Analysis comparable to the one described in the preceding section was performed for ACTpol, and that analysis suggested an acceptably low level of sidelobes from the cold optics tube. However, in-situ near-field measurements of the ACTpol optics tubes show a much higher level of scattered and/or diffracted light at large angles \cite{gallardo2018far}, possibly due to an excess reflectivity in baffling materials inside the camera. In this section, we take this measured optics tube spillover as a worst case scenario and use it to inform the design of an additional baffle that can be installed inside the SO elevation structure. 

\begin{figure}[]
    \centering
    \includegraphics[width=0.48\textwidth]{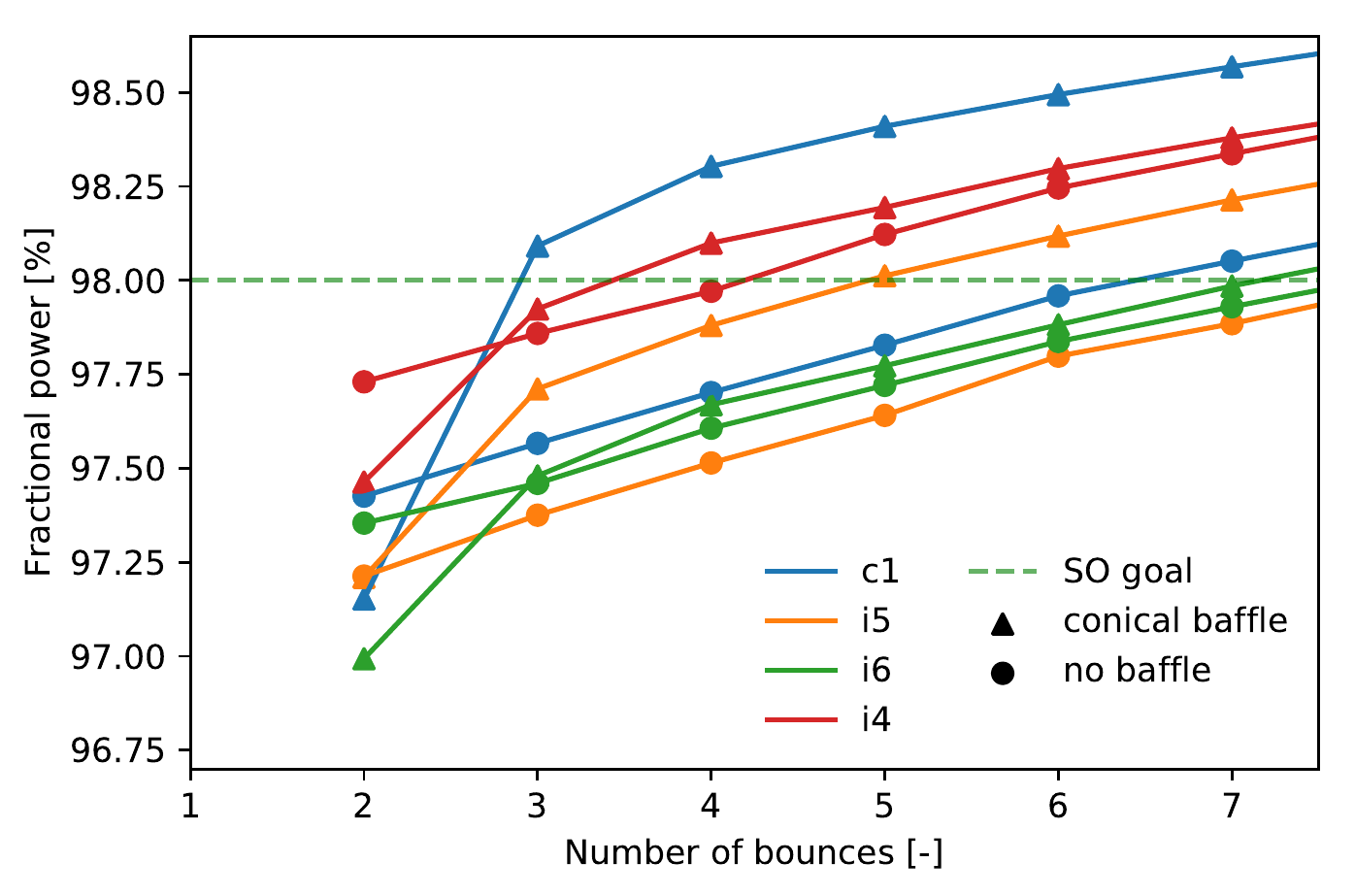}
    \caption{Fractional optical power (compared to input) that makes it out of the elevation housing aperture as a function of the number of bounces for a time-reverse optical ray trace assuming a camera beam distribution as measured in \cite{gallardo2018far}. The four different colors correspond to the four optics tubes considered, \textsf{c1, i4, i5} and \textsf{i6}. Triangles show the power transmitted to the sky from a reflective conical baffle, while circles show power at sky from using no baffle at all. For the configuration that includes the baffle, the jump in power transmitted to the sky between 2 and 3 bounces primarily comes from rays bouncing off the receiver baffle to the secondary then the primary and then to the sky. The horizontal dashed green line, labeled "SO goal", is part of a suite of instrument performance metrics that are used in discussion of SO mapping speed \citep{SO2019}.}
    \label{fig:power_on_sky}
\end{figure}

\begin{figure*}[t!]
    \centering
    
    \begin{minipage}{0.45\textwidth}    
        \includegraphics[width=\textwidth]{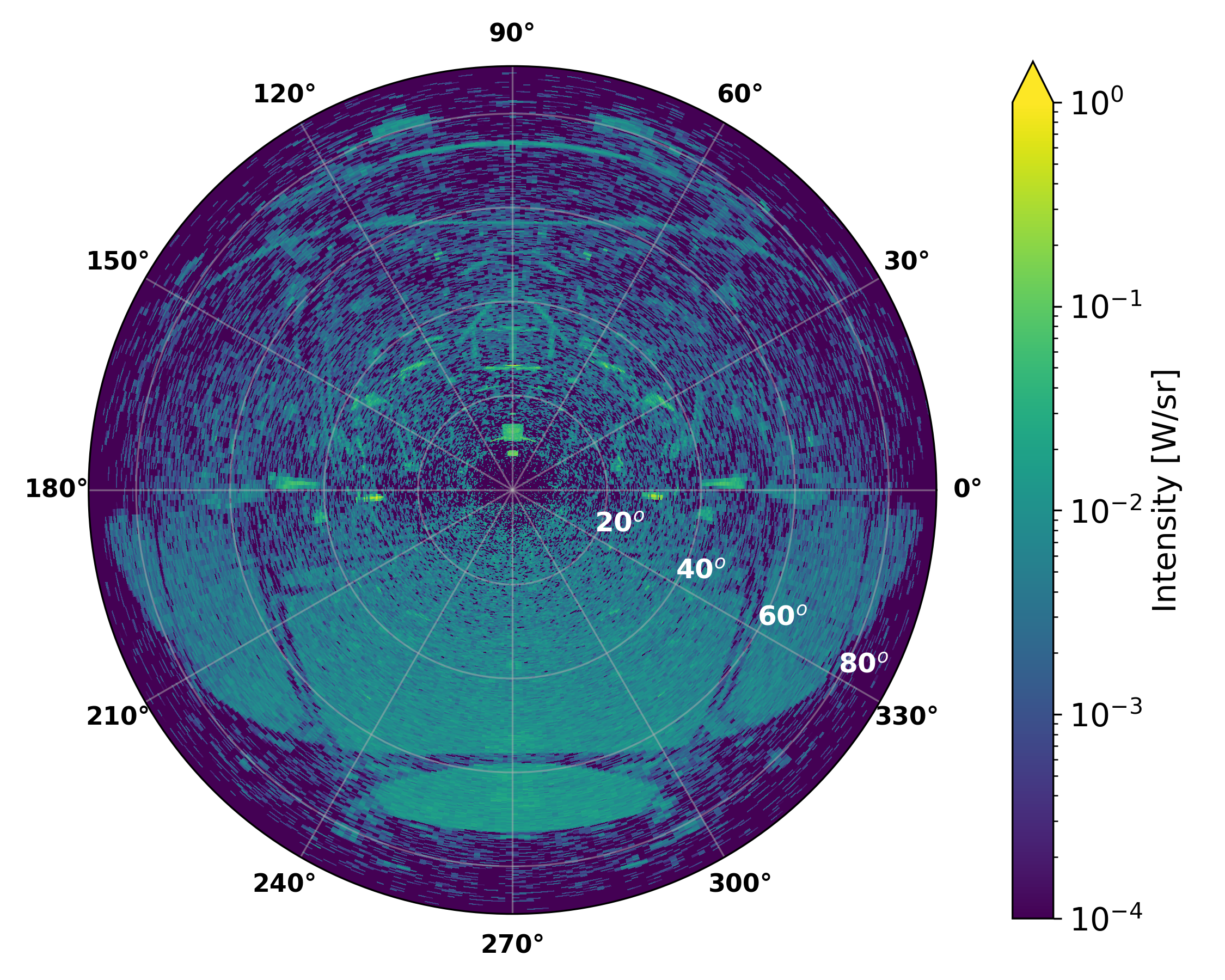}
    \end{minipage}
    \begin{minipage}{0.45\textwidth}    
        \includegraphics[width=\textwidth]{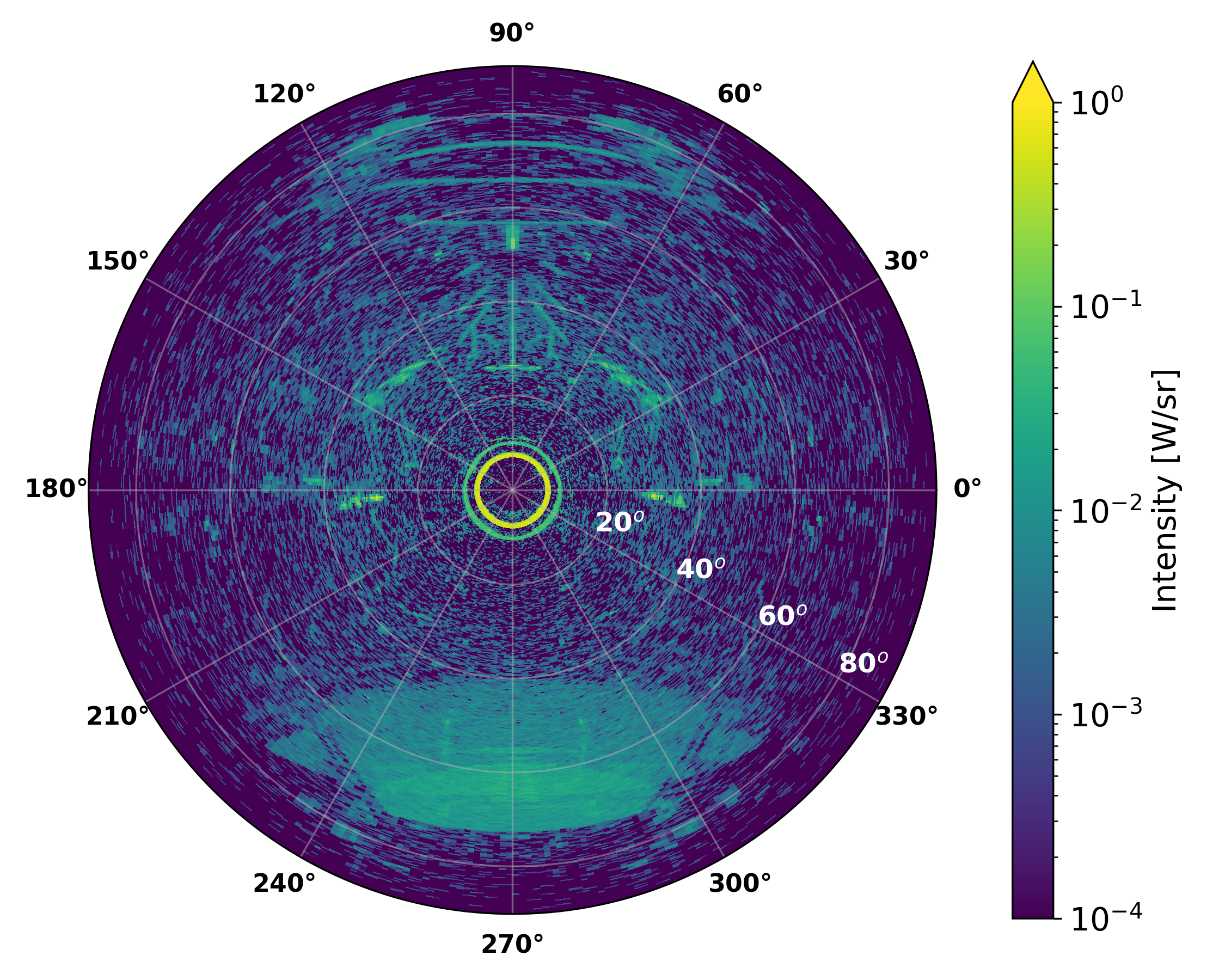}
    \end{minipage}
        
\caption{Left: LAT spillover pattern (at five bounces) for center optics tube achieved with a reflective cabin and no forebaffles attached to the receiver entrance. Total power integrates to approximately \SI{0.98}{\watt} instead of \SI{1.0}{\watt} due to the limit in the number of bounces and the $\sim 2\%$ loss imposed to try to capture effects from non-optical component in the system. The center of the figure corresponds to the location of the main beam. The line from 90 to 270 degrees (in azimuth) corresponds to the elevation axis in this projection. The large sidelobe at \ang{270} in the azimuthal coordinate corresponds to the direct line of sight from the camera to the sky (see Fig.\ \ref{fig:LAT_3D}). The symmetric features seen roughly \ang{30} from the main beam centroid at \ang{0} and \ang{180} azimuthal angle, correspond to rays that have bounced four times, twice off the inside walls of the elevation housing and twice off the reflectors. Right: Corresponding sidelobe pattern when a reflective conical baffle is attached at the camera entrance. Note that the extended region centered on \ang{270} is reduced significantly in both size and power (the long tails at \ang{210} and \ang{330} in azimuth point to the ground as the telescope points away from the zenith). Approximately \SI{0.9}{\percent} of the total optical power resides in an annulus roughly 7-\ang{8} from the main beam.}
\label{fig:sidelobes_prediction}
\end{figure*}


During the design stages of SO LAT, we studied geometrical mitigation strategies to (in a time-reverse sense) direct as much radiation from the optics tubes to the sky as quickly as possible (i.e., with the lowest number of bounces). We use a source with a beam given by an analytic model, placed at the secondary focus and oriented parallel to the chief ray (see Fig.\ \ref{fig:LAT_3D_spill}). This beam causes roughly \SI{3}{\percent} of the total optical power to spill past the secondary mirror (see Eq.~1 and Fig.~4 in \cite{gallardo2018far}). The model is purely phenomenological and does not impose restrictions on the physical origin of spilled light. The model is composed of a central Gaussian beam and an exponential fall-off tail (for the analytic description, see Table 1 in \cite{gallardo2018far}). A non-sequential time-reverse ray trace that incorporates a preliminary 3D solid model of the warm optics (see Fig.~\ref{fig:LAT_3D_spill}) is used to compute the sky illumination pattern caused by interactions with the telescope structure. We treat all surfaces in the elevation structure as reflectors with a $\sim$\SI{2}{\percent} loss. This simulation was implemented in Zemax in non-sequential mode and only considers unpolarized light; we leave polarization as future work. Further details of this ray tracing analysis will be presented in a future work.



The optimization of the warm structure was done in successive iterations as the mechanical design converged. We tested the benefit of using guard rings around the warm mirrors and found that the benefit would be marginal compared to the cost. Despite this, we installed mounting points to allow a later addition if deemed necessary. From this analysis, we concluded that apertures inside the elevation structure (like the elevation bearing or the entrance aperture on the top of the structure) should be made as large as possible to allow light to escape the housing with a minimum number of reflections. 

We also implemented a receiver baffle surrounding the camera entrance apertures to direct stray light from the receiver to the sky. We considered two design options for this purpose, a truncated paraboloid with a focal point at the midpoint of the receiver entrance and a simpler conical shape. The paraboloid configuration was shown to perform best for rays placed at the center of the secondary focal plane, but after accounting for proper geometrical weighting of the entire secondary focal plane, the simpler conical shape performs better. The conical design is also less expensive and more mechanically robust. The conical baffle (shown in Fig.~\ref{fig:LAT_3D_spill}) covers all the available space in the receiver cabin to smoothly transition the aperture size to the elevation housing. 



Figure~\ref{fig:power_on_sky} shows the fraction of optical power that makes it to the sky as a function of the number of ray tracing bounces. For all field locations considered, the nominal path (2 bounces) gets 97.0--\SI{97.8}{\percent} of the power to the sky. For the model that includes the receiver baffle, there is a significant increase seen in the amount of power that makes it to the sky between 2 and 3 jumps. It is estimated that this baffle can boost the fraction of light making it to the sky by as much as 0.5\si{\percent} of the total input optical power (depending on the field location), which would have an impact of roughly 10-\SI{15}{\percent} in mapping speed at MF frequencies \cite{Hill2018}, at the expense of having a circularly symmetric sidelobe at 7--8 degrees from the main beam.

Figure~\ref{fig:sidelobes_prediction} summarizes the results of our ray tracing simulations for the conical baffle. In general, this conical baffle design suppresses very large angle sidelobes, giving them a more central and azimuthally symmetric distribution on the sky. Of particular interest is an extended feature centered around the \ang{270} azimuthal coordinate. This feature is due to direct illumination of the elevation housing aperture from the source location at the receiver vacuum windows. For the case of the central field location, which is shown in Fig.~\ref{fig:sidelobes_prediction}, we find that the amount of power in this large sidelobe pattern is reduced by a factor of 1.6, going from 1.00 to \SI{0.63}{\percent} of the total beam power. The extended sidelobe is replaced by an annular feature that spans roughly 7--\ang{8} and contains approximately \SI{0.9}{\percent} of the total power that makes it to the sky. If this model is accurate, we will be able to measure, avoid in observation-time and correct for this feature in analysis. It is also worth emphasizing that the assumption of \SI{3}{\percent} secondary spillover is likely pessimistic given the research and development on internal baffling \cite{Chesmore2020}. 





%% file: 5PhysicalOptics/po.tex
\section{Physical optics}
\label{sec:po}

\begin{figure}[t!]
    \centering
    \includegraphics[width=0.45\textwidth]{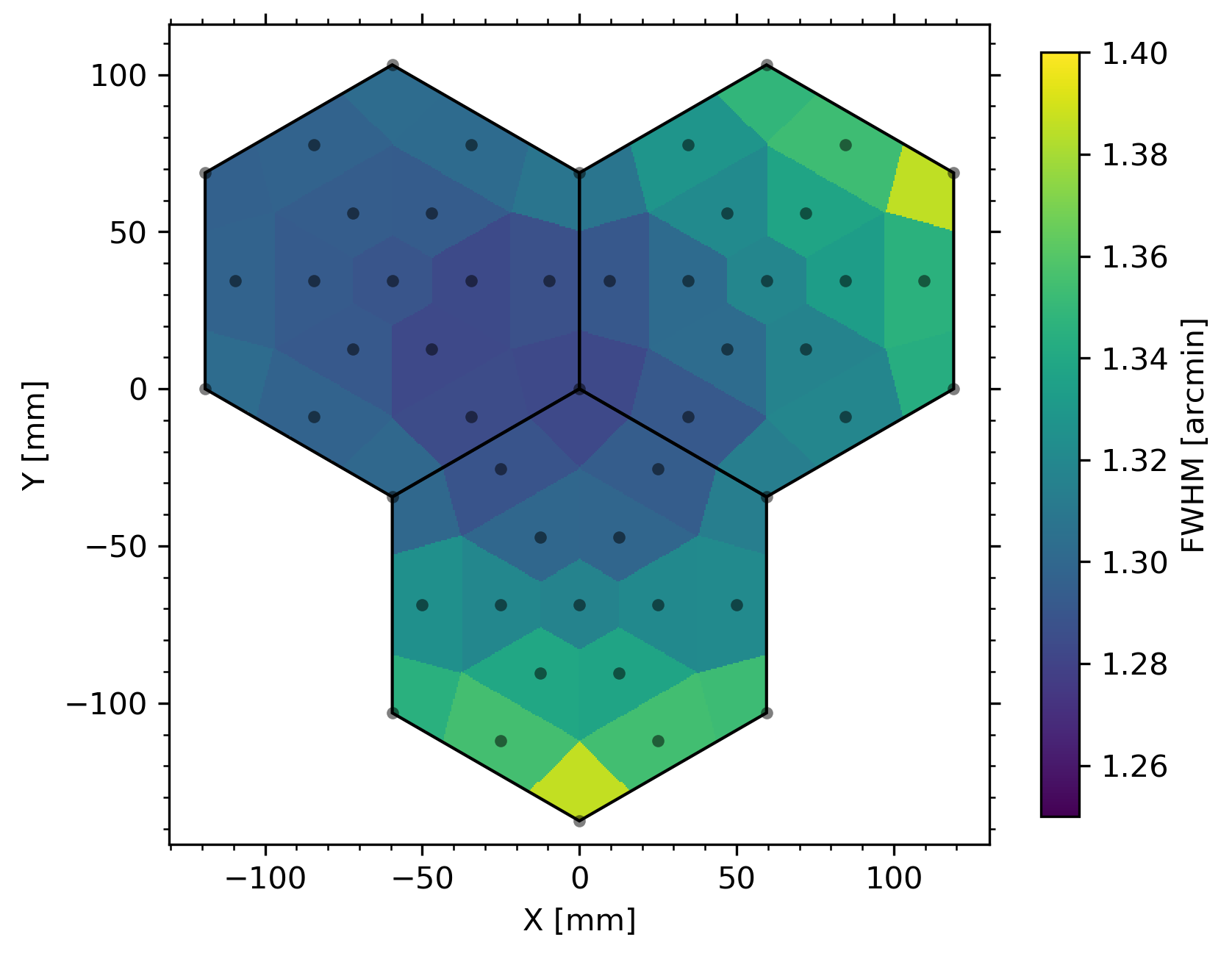}    
    \caption{The distribution of predicted beam FWHM at \SI{150}{\giga\hertz} across the \textsf{i4} optics tube (see Fig.~\ref{fig:LAT_label}). The markers represent the location of the simulated pixels. Note that there are markers on all hexagon vertices. For the purposes of highlighting coverage, the plotted colormap is based on nearest-neighbor interpolation. Note that statistics presented in other parts of the paper are based on bi-linear interpolation.}
    \label{fig:tube8_fwhm_150}
\end{figure}

\subsection{Basic model setup}
Physical optics simulations allow us to calculate the frequency dependent beam response, including diffraction effects, in both the near- and far-field. Using this technique, we can predict the nominal beam response for the different frequency bands in the SO LAT. W
e generate an ensemble of physical optics simulations using an application programming interface (API) that couples to TICRA Tools (formerly GRASP) \cite{TicraTools}; a similar approach was described in \cite{Duivenvoorden2018, Gudmundsson2020}.\footnote{See \url{https://www.ticra.com/}} The physical optics simulations are carried out in a time-reverse sense (transmit) and start with the generation of outward propagating electric field at the location of the focal plane. The field is then propagated through all three lenses of each optics tube as well as the cold stop. The field from lens 1 is then used to calculate the field distribution on the hexagonal vacuum window (see Fig.~\ref{fig:cold_optics}) which is in turn used to predict the field at the secondary and primary mirrors and then into the far-field. 

The optically active region of the secondary and primary mirrors are input to these simulations in the form of a tabulated mesh sampled at 1-cm intervals. Within TICRA Tools, the regions are defined as ellipses with 6003- and \SI[number-unit-product=\text{-}]{6344}{\milli\metre} geometrical mean diameters for the secondary and primary, respectively. This implies that the TICRA tools models extend beyond both the optically active area of the mirrors as defined by geometrical optics and the as-built reflectors which are composed of a total of 146 rectangular panels (see Fig.~\ref{fig:LAT_3D}). As a result, the predicted far-field beam maps will not capture diffraction effects from truncated field distributions caused by the discrete panel distributions; this will likely affect the predicted beam size at the percent-level. The tabulated mesh of the elliptical reflector shapes are resampled by TICRA Tools at the lowest resolution required to be accurate down to \SI{-40}{\deci\bel} compared to peak main beam response. Similarly, the surfaces on each side of the three lenses as well as the hexagonal vacuum window are meshed at the resolution that is sufficient for convergence of the far-field beam maps.

\begin{figure*}[t!]
    \centering
    \includegraphics[width=0.48\textwidth]{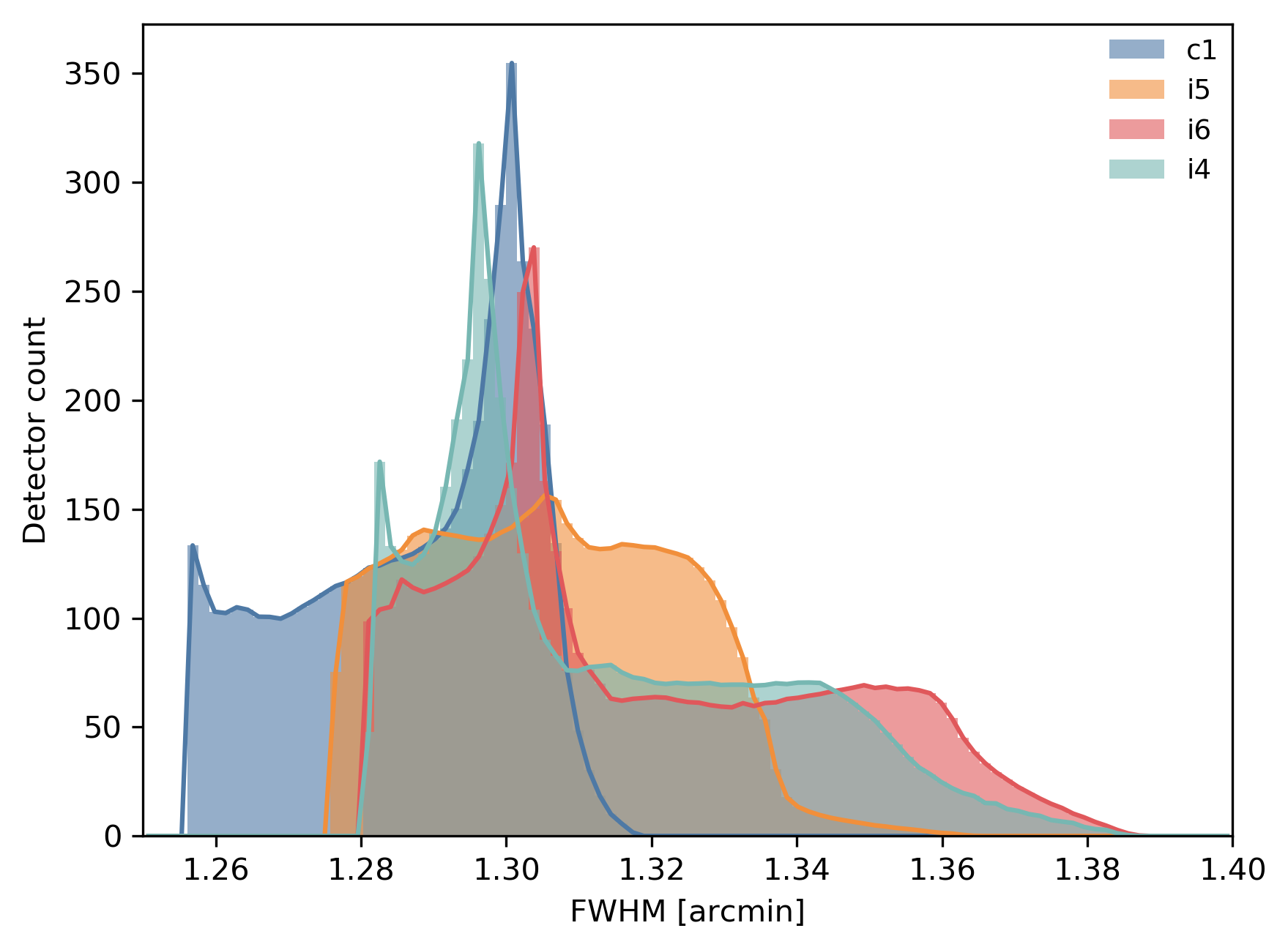}
    \includegraphics[width=0.48\textwidth]{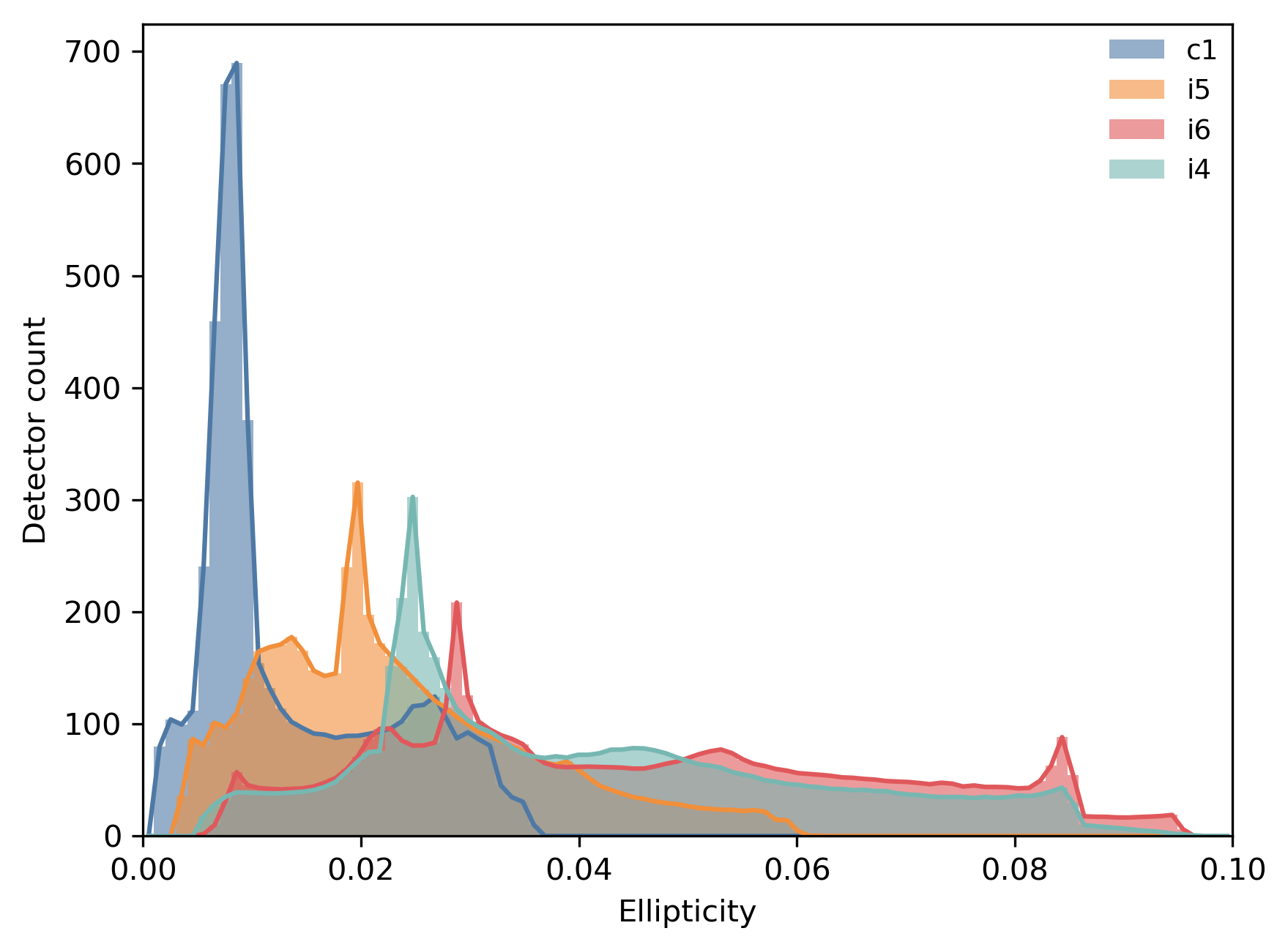}
    \caption{\textit{Left}: Distribution of the predicted 150-GHz beam FWHM for four LAT optics tubes based on physical optics simulations. As expected, the beams of the center optics tube are most compact while tubes \textsf{i6} and \textsf{i4} perform the worst. \textit{Right}: Distribution of the predicted 150-GHz beam ellipticities for four LAT optics tubes based on physical optics simulations. As expected, the beams of the center optics tube are most symmetrical while tubes \textsf{i6} and \textsf{i4} are more elliptical.}
    \label{fig:fwhm_dist_150}
\end{figure*}

The physical optics simulations do not account for internal reflections in the lenses or any back scattering because of limitations in simulations of anti-reflection coatings. Similarly, the simulations do not account for interactions with absorbing elements along the sides of the cooled optics tubes nor do they account for in-band scattering from filters or (lens/reflector) surface irregularities. Finally, effects from interactions with the telescope housing structure or its aperture is not included in the physical optics simulations described below. The two reflectors incorporated in these physical optics simulations are a single-body object with an elliptical rim; the effects of the 146 panels that make up the mirrors, as well as the gaps between them, are not included in these simulations. All of these simplifying assumptions are made to facilitate practical computation times. With this in mind, it is clear that the simulation results describe theoretical best-case scenarios given the proposed telescope geometry and predicted detector far-field beam. We have not conducted tolerancing analysis using physical optics, therefore we do not have means to quantify the expected deviations from mechanical errors and deformations in optical elements. However, past experience from the Atacama Cosmology Telescope (ACT) indicates that for the MF frequency bands we can expect a 5-\SI{10}{\percent} increase in realized beam FWHM compared to naive $\lambda / D$ predictions \cite{Choi2020}. From simply comparing the Strehl ratio distributions for ACT and SO LAT at MF frequencies, we would expect SO LAT to perform somewhat better. 

We note that it is possible to run physical optics simulations that account for the finite conductivity of the aluminum panels. For a select number of detectors we have run simulations that assume a finite conductivity of $\sigma_\mrm{Al} = \SI{2.5e7}{\siemens\per\meter}$ for both the secondary and primary mirrors. However, we find that the impact on both the estimated beam FWHM and ellipticity is negligible, and therefore we have not included it in our general simulations as it adds significantly to the computation time.

During nominal operations, the scan-averaged telescope elevation is approximately \ang{45}. As the telescope changes elevation angles, the LATR will co-rotate so as to keep the illumination pattern of the optics constant on the secondary. The clocking of the LATR will also be configured for optimal cryogenic performance. This implies an orientation in which the LATR dilution refrigerator is oriented at 12 o'clock (pointed up) as seen from the center of the secondary mirror while the telescope is pointed at \ang{45} elevation. If the elevation housing rotates while the LATR is fixed, we expect some changes in the beam response. The physical optics simulations described in this section correspond to a LATR orientation where the dilution refrigerator has been rotated \ang{30} from its nominal configuration. This happens to be the configuration that yields the highest band-average Strehl numbers. We have performed PO simulations to compare the expected far field beams for these two configurations. Basic far field beam properties such as FWHM and ellipticities are found to be consistent within 1-\SI{2}{\percent}. Initial optimization of the LAT scan strategy can be found in \cite{Stevens2018}. Further details will  be described in future publications. 

To map out beam properties across the focal plane, we run physical optics simulations where the source is placed at 52 distinct FPU locations within each optics tube. Because of symmetry, we only consider tubes \textsf{c}, \textsf{i5}, \textsf{i6}, and \textsf{i4} (see Fig.~\ref{fig:LAT_3D}). The results of these simulations are analyzed to estimate beam properties such as beam width, ellipticity, and cross-polar response. These properties are then input to a bi-linear interpolation routine, {\tt scipy.interpolate.griddata}, to estimate the distribution across the entire focal plane (see the following sub-sections). 

\subsection{Pixel beam model}
The MF receivers will employ spline-profiled feedhorn arrays which are machined out of aluminum and then gold plated \citep{Granet2004, Simon2018}. A waveguide section of the feedhorns then couples the incoming radiation to bolometer arrays \cite{Hubmayr2018}. The feedhorns have \SI[number-unit-product=\text{-}]{5.15}{\milli\meter} apertures and \SI{0.15}{\milli\meter} wall thickness between feeds. These produce roughly $-1.3$ and \SI{-2.9}{\deci\bel} cold stop edge tapers at 90 and \SI{150}{\giga\hertz}, respectively. Simulations of feedhorn optical response are performed using Ansys HFSS. During the feedhorn design process, requirements of the far-field pixel beam symmetry was relaxed in order to increase optical throughput. The output of the HFSS simulations is used as input to the TICRA Tools physical optics simulations. For simplicity, the input pixel beams are perfectly polarized so that any cross-polar far-field response will be indicative of optical effects and non-idealities outside of the pixels. The simulations results represent the monochromatic beam at the specified frequencies. 

\subsection{Beam size and ellipticity}
Beam size and symmetry impact the science goals of the LAT. We use the physical optics simulation results to estimate these properties for the telescope under nearly ideal circumstances. The results from these simulations are summarized in Fig.~\ref{fig:tube8_fwhm_150}-\ref{fig:fwhm_dist_150} and Table~\ref{tab:po_beam_prop}. As expected, both the beam size and ellipticity grow as one moves away from the center optics tube. We define the beam full width at half maximum (FWHM) as the geometrical mean of the FWHM for the major and minor axis of the best fit elliptical Gaussian model. The beam ellipticity is defined as 
\begin{equation}
e = \frac{\sigma_{\mrm{max}}-\sigma_{\mrm{min}}}{\sigma_{\mrm{max}}+\sigma_{\mrm{min}}},
\end{equation}
where $\sigma _{\mrm{max}}$ and $\sigma _{\mrm{min}}$ correspond to the widths of the best-fit Gaussian envelope to the major and minor axis of the co-polarized far-field beam response. The beam FWHM, $\theta _\mrm{FWHM}$, is related to $\sigma_\mrm{max}$ and $\sigma_\mrm{min}$ according to:
\begin{equation}
\theta _\mrm{FWHM} = \sqrt{8 \ln (2) \sigma_\mrm{max} \sigma_\mrm{min}}.
\end{equation}
Assuming a \SI[number-unit-product=\text{-}]{5.5}{\metre} primary aperture illumination, the beam sizes are predicted to stay within roughly 5-\SI{10}{\percent} of the diffraction limit at 90 and \SI{150}{\giga\hertz}, respectively (see Table \ref{tab:po_beam_prop}). The deviation from the theoretical best-case scenario is primarily due to non-uniform aperture illumination. As expected, the center optics tube shows the best optical performance and there is strong correlation between beam size and Strehl ratio.

\begin{figure*}[t!]
    \centering
    \includegraphics[width=0.48\textwidth]{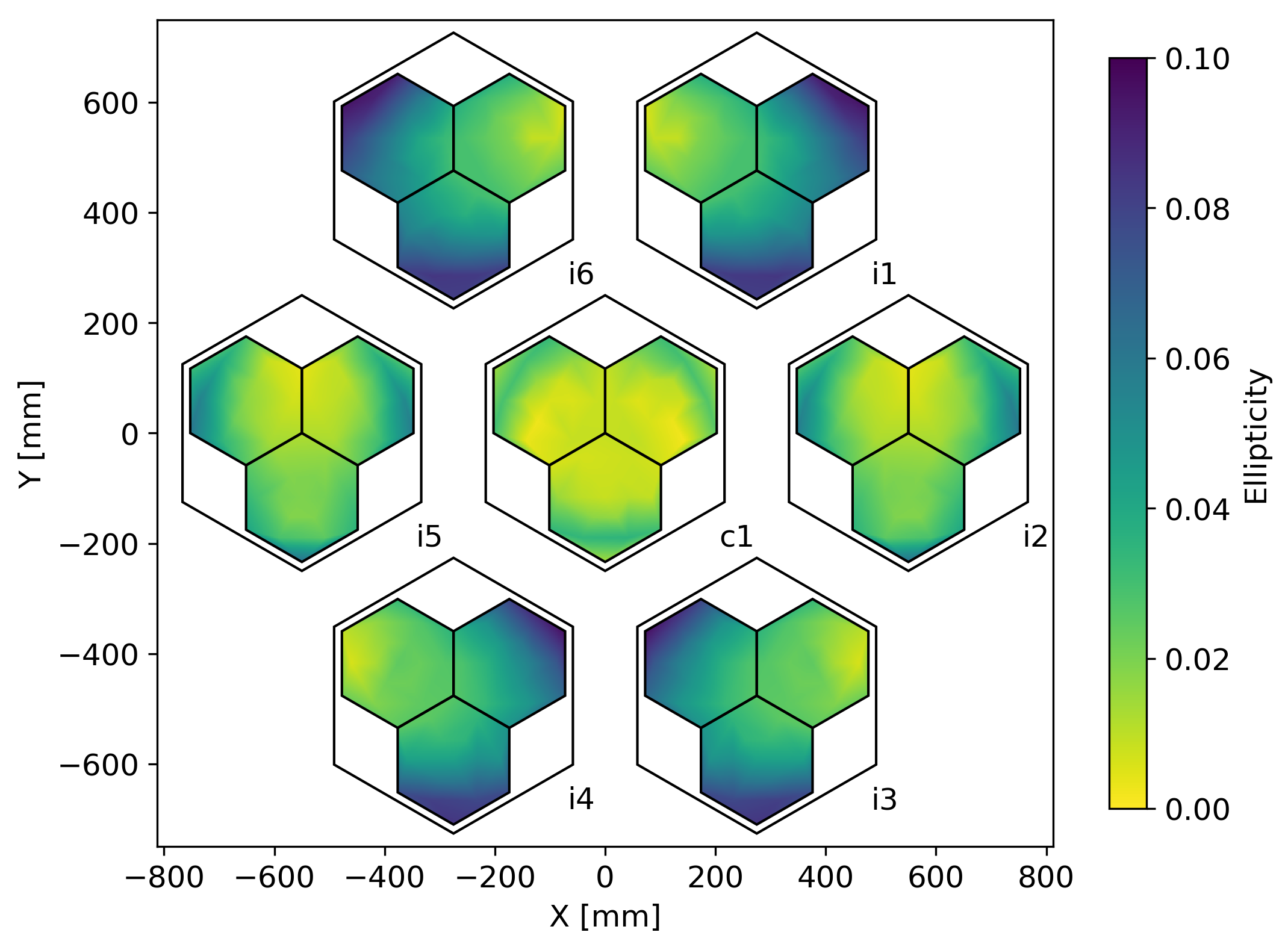}
    \includegraphics[width=0.48\textwidth]{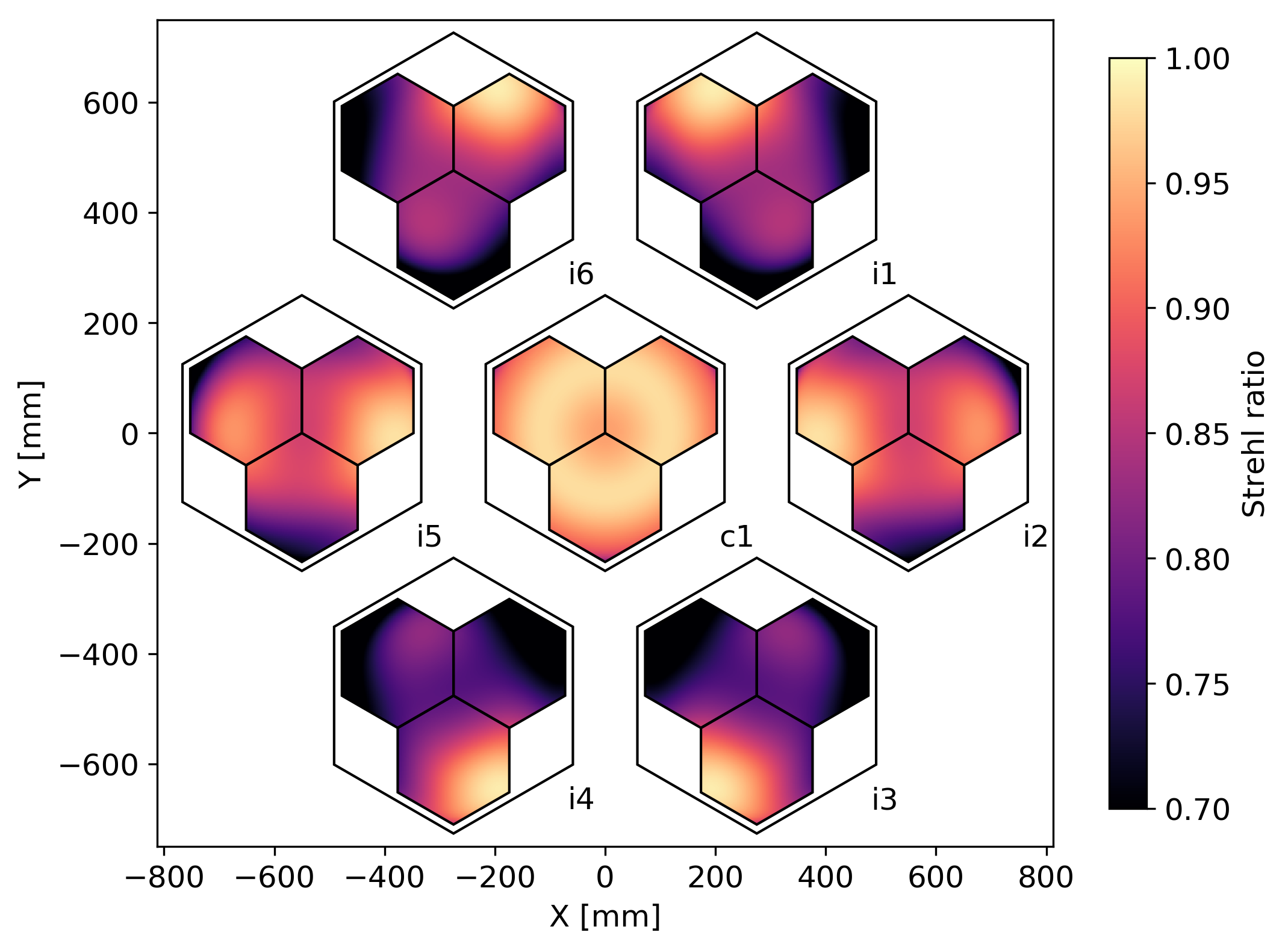}
    \caption{\textit{Left}: Distribution of predicted beam \SI{150}{\giga\hertz} ellipticity across the different telescope tubes as seen when looking (along the optical axis) from the secondary mirror towards the LATR (see Fig.~\ref{fig:LAT_label}). Note that the ellipticity has a non-trivial dependence on distance from the system symmetry axis. The large hexagon represents the outline of the hexagonal vacuum window while the 3-tile configuration represents the focal plane after applying a factor of 1.7 scaling for clarity. \textit{Right}: Distribution of Strehl ratios at \SI{270}{\giga\hertz} (from Zemax) as a function of focal plane location. Note that this graph represents the Strehl ratio as a function of physical location on the focal plane, not angle on the sky. The 90 and \SI{150}{\giga\hertz} Strehl ratios can be trivially calculated from the \SI{270}{\giga\hertz} Strehl ratio using the wavelength ratio, but we choose to show the latter because it demonstrates where the optical performance dips below the so-called diffraction limit which corresponds to Strehl ratio of 0.8.}
    \label{fig:beam_dist}
\end{figure*}

For the most part, beam ellipticity is found to correlate strongly with Strehl ratio as calculated using Zemax. This of course implies that the beams get more elliptical (on average) with increasing frequency. However, since the Strehl ratio statistics assume uniform aperture sampling instead of applying weights that represent the profile of the feedhorns, we do not expect perfect correlation between the beams predicted by these physical optics simulations and the quoted Strehl ratios. Fig.~\ref{fig:beam_dist} shows the distribution of \SI[number-unit-product=\text{-}]{150}{\giga\hertz} beam ellipticity across the seven optics tubes. For comparison, the \SI[number-unit-product=\text{-}]{270}{\giga\hertz} Strehl ratio distributions are also shown. From these simulations at 90 and \SI{150}{\giga\hertz}, we observe that the orientation of the semi-minor and semi-major axis of the best-fit elliptical beam is strongly correlated with the projected aperture of the primary and secondary mirrors and less dependent on the illumination profile of the cold stop.



 \subsection{Cross-polarization}
 Cross-polar response is known to be minimal for Crossed-Dragone telescope designs \citep{Mizuguchi1978, Dragone1978}. However, because of off-axis pixels (which break the Mizuguchi-Dragone condition), finite reflector conductivity, and other non-idealities, some level of cross-polar response is unavoidable. The physical optics simulations allow us to estimate cross-polar response caused by geometrical effects and finite conductivity. We calculate the far-field Stokes $I$, $Q$, and $U$ beams from the complex electric fields output by the TICRA Tools simulations. Since the pixel beam input to our simulation is perfectly polarized, the Stokes $Q$ beam is very similar to the Stokes $I$ beam. The $U/I$ beam map, on the other hand, has a typical quadrupolar shape with alternating positive and negative lobes. After accounting for polarization angle rotation, so as to minimize the solid angle in the $U$ beam, we calculate the peak ratio of $U/I$ at the map level. This number gives an indication of the level of cross-polarization that is expected from the optical system alone. Note that this does not incorporate the cross-polar response from the feedhorns or from inductive coupling in the readout electronics. At 90 and \SI{150}{\giga\hertz}, we find that the peak amplitude in the $U/I$ beam map ranges from $-30$ to \SI{-20}{\deci\bel} with the lowest values registering at the center of each focal plane. 

\begin{table}
\caption{Telescope-averaged beam properties at 90, 150, 220, and \SI{270}{\giga\hertz}. Numbers represent median and lower/upper limits spanning 15.9-84.2th percentile range of the distributions obtained from interpolating the results of 52 physical optics simulations distributed across the entire focal plane (see Fig.\ \ref{fig:beam_dist}). We note that the expected FWHM from simply assuming  $\lambda / D$, with an effective diameter $D = \SI{5.5}{\meter}$, gives 2.03, 1.22, 0.83, and \SI{0.68}{\arcmin} for 90, 150, 220, and \SI{270}{\giga\hertz}, respectively. At 220 and \SI{270}{\giga\hertz} we observe larger deviations from the naive $\lambda / D$ relation because the impact of optical non-idealities becomes stronger at higher frequencies.}
\label{tab:po_beam_prop}
\begin{tabular}{llll}
Freq.\ & Tube & FWHM & Ellipticity  \\
  \lbrack GHz \rbrack &  & [arcmin] & [-]  \\
\hline
90 & \textsf{c} & $2.072^{+0.017}_{-0.024}$ & $0.004^{+0.002}_{-0.002}$ \\
90 & \textsf{i5} & $2.093^{+0.018}_{-0.023}$ & $0.007^{+0.004}_{-0.004}$ \\
90 & \textsf{i6} & $2.097^{+0.022}_{-0.025}$ & $0.016^{+0.007}_{-0.006}$ \\
90 & \textsf{i4} & $2.093^{+0.021}_{-0.024}$ & $0.014^{+0.006}_{-0.005}$ \\
150 & \textsf{c} & $1.290^{+0.015}_{-0.019}$ & $0.009^{+0.012}_{-0.007}$ \\
150 & \textsf{i5} & $1.305^{+0.019}_{-0.020}$ & $0.021^{+0.013}_{-0.012}$ \\
150 & \textsf{i6} & $1.309^{+0.032}_{-0.026}$ & $0.043^{+0.027}_{-0.023}$ \\
150 & \textsf{i4} & $1.304^{+0.028}_{-0.022}$ & $0.036^{+0.024}_{-0.018}$ \\
220 & \textsf{c} & $0.946^{+0.018}_{-0.020}$ & $0.026^{+0.028}_{-0.021}$ \\
220 & \textsf{i5} & $0.962^{+0.030}_{-0.029}$ & $0.046^{+0.029}_{-0.030}$ \\
270 & \textsf{c} & $0.832^{+0.023}_{-0.027}$ & $0.040^{+0.042}_{-0.034}$ \\
270 & \textsf{i5} & $0.846^{+0.042}_{-0.042}$ & $0.070^{+0.047}_{-0.047}$ \\
\hline
\end{tabular}
\end{table}



\subsection{Cold spillover results}
With the pixel beam models as input, the physical optics simulations can be used to keep track of how much power spills past a given optical element. These calculations therefore allow us to estimate the total optical spillover inside the optics tube as well as the spillover past the two reflectors. As discussed in Sections \ref{sec:go} and \ref{sec:spillover}, spillover influences the optical performance and mapping speed of the telescope.

\begin{figure}[]
    \centering
    \includegraphics[width=0.48\textwidth]{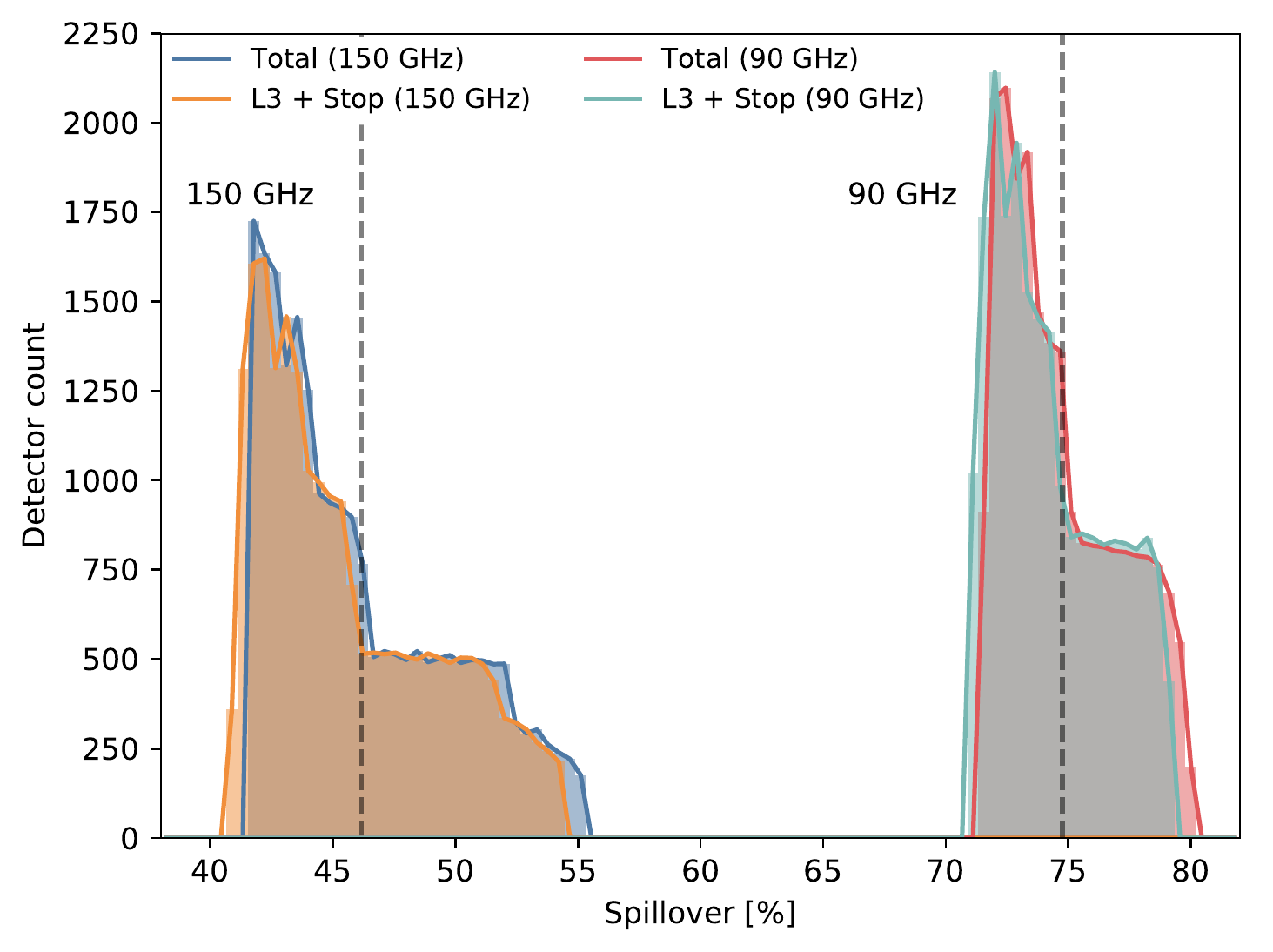}
    \caption{Cold spillover at 90 and 150 GHz as a fraction of the total optical power. Blue and red curves trace the outline of the total estimated cold spillover while the orange and teal curves represent the power that spills past either lens 3 (the one closest to the focal plane) or the cold stop. Vertical dashed lines represent the mean of the total spillover distributions.}
    \label{fig:spillover}
\end{figure}

The non-sequential ray tracing enables fast mitigation of spurious rays and therefore sheds light on both systematic and noise performance; this approach is also the only viable option for electrically large objects like the LAT telescope housing. On the other hand, a physical optics analysis quantifies nominal spillover performance expected from a perfect realization of a proposed design. Using a programming interface to TICRA Tools, we can run spillover analysis as a function of focal plane location. This approach also expands and improves on the Gaussian beam formalism that is frequently used (see e.g. \cite{Griffin2002}).

Unfortunately, the physical optics spillover calculations require a higher degree of convergence and therefore higher resolution meshing compared to those needed to estimate the far-field beam response. Because these are more time consuming, we are not able to run spillover simulations for dozens of detectors in each of the different optics tubes. We also choose not to focus on these reflector spillover statistics, since near field sidelobe estimates, for example from in-band scattering on filters or internal reflection on both active and passive optical elements inside the tube, have considerable uncertainties. In order to generate spillover estimates at a given frequency, we have run simulations at only seven points along a radial line from the center of the focal plane to the edge. Internally to the optics tubes, the total spillover past a given optical element is azimuthally symmetric. From these seven physical optics simulations we can estimate the spillover at any given location on the focal plane through interpolation.


Figure \ref{fig:spillover} shows a histogram of total optical spillover internal to the tube. The steps are largely caused by the radial pixel distributions for the three hexagonal tile arrangement. At \SI{150}{\giga\hertz} we note that the cold spillover can vary by up to \SI{40}{\percent} depending on focal plane location, but only by about \SI{15}{\percent} at \SI{90}{\giga\hertz}. As expected, the majority of the cold spillover happens at or before the cold stop. However, percent-level spillover past lens~1 and lens~2 is also observed in some cases, which is manifest in the slight shift of the total spillover distributions relative to the ``L3 + Stop'' distribution. It is interesting to note that approximately \SI{75}{\percent} of the \SI[number-unit-product=\text{-}]{90}{\giga\hertz} pixel beam spills on the walls of the cryogenic receiver (past the stop).




\subsection{Frequency and focal plane statistics}
Some basic far-field beam statistics as a function of frequency and focal plane location are shown in Table \ref{tab:po_beam_prop}. As is clear from Fig.\ \ref{fig:beam_dist}, the distribution of the beam properties are not Gaussian; this is largely due to the geometrical sampling of the focal plane from the 3-tile arrangement.

%% file: 6PanelGaps/6panelgaps.tex
\section{Panel gaps effect}
\label{sec:panels}

\begin{figure}[t!]
    \centering
    \includegraphics[width=0.5\textwidth]{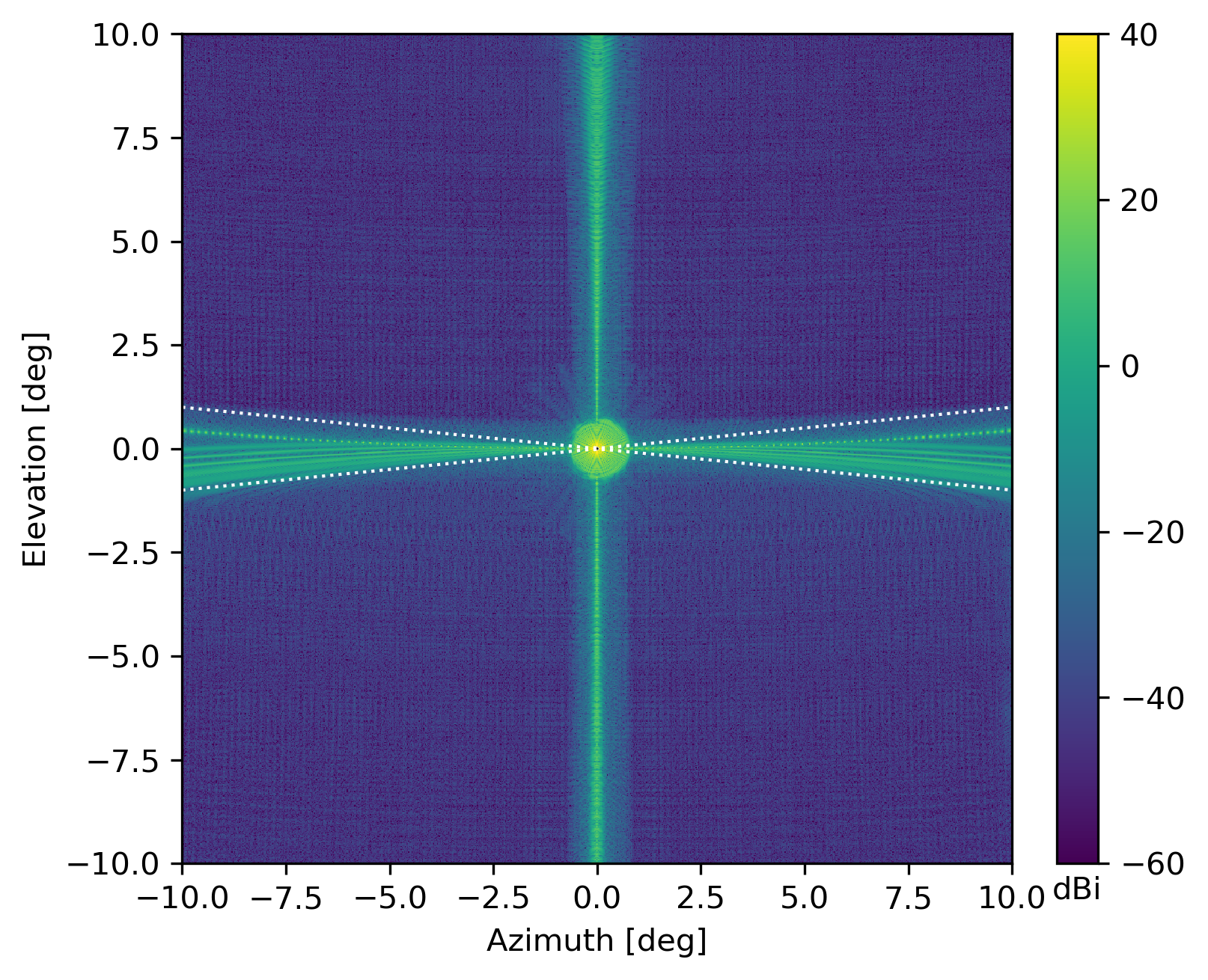}
    \caption{Co-polar far-field beam map for a 150-GHz detector located in the center optics tube with panel gap separation at \SI{1.2}{\milli\meter}. Cross-shaped sidelobe patterns are clearly visible at roughly \SI{20}{\deci\beli}. This cross-shaped pattern is not seen in simulations that treat the reflectors as a gapless reflectors. The forward gain for this detector is roughly \SI{78}{\deci\beli}. Dotted white lines bound the region used to calculate beam profiles for Fig.~\ref{fig:beam_profile}.}
    \label{fig:panel_gap_beam_map}
\end{figure}

The primary and secondary mirrors are composed of 69 and 77 rectangular panels, respectively. The primary panels are $750\times670$ mm, while the secondary are $700\times710$ mm. They are arranged in a square array, with a nominal gap of \SI{1.2}{\milli\meter} between adjacent panels when the mirrors are at a temperature of \SI{-6}{\celsius}. These discontinuities in the reflector surface are expected to induce features in the far-field beam patterns. To quantify this, we employed a TICRA Tools physical optics analysis method which defines the reflectors as multi-face objects. Each panel is individually defined by a rim placed on the tabulated mesh described in Section~\ref{sec:po}. The far-field beam map is computed on a $10 \times 10$ degree elevation-over-azimuth grid. The source that illuminates the secondary is placed at the center of the central optical tube. This source illuminates the secondary mirror using a 90- or a \SI[number-unit-product=\text{-}]{150}{\giga\hertz} near-field beam pattern which is based on the PO model of the receivers described in Section \ref{sec:po}. This is done to simulate the tapered illumination at the nominal rim of the mirrors.

\begin{figure}[t!]
    \centering
    \includegraphics[width=0.5\textwidth]{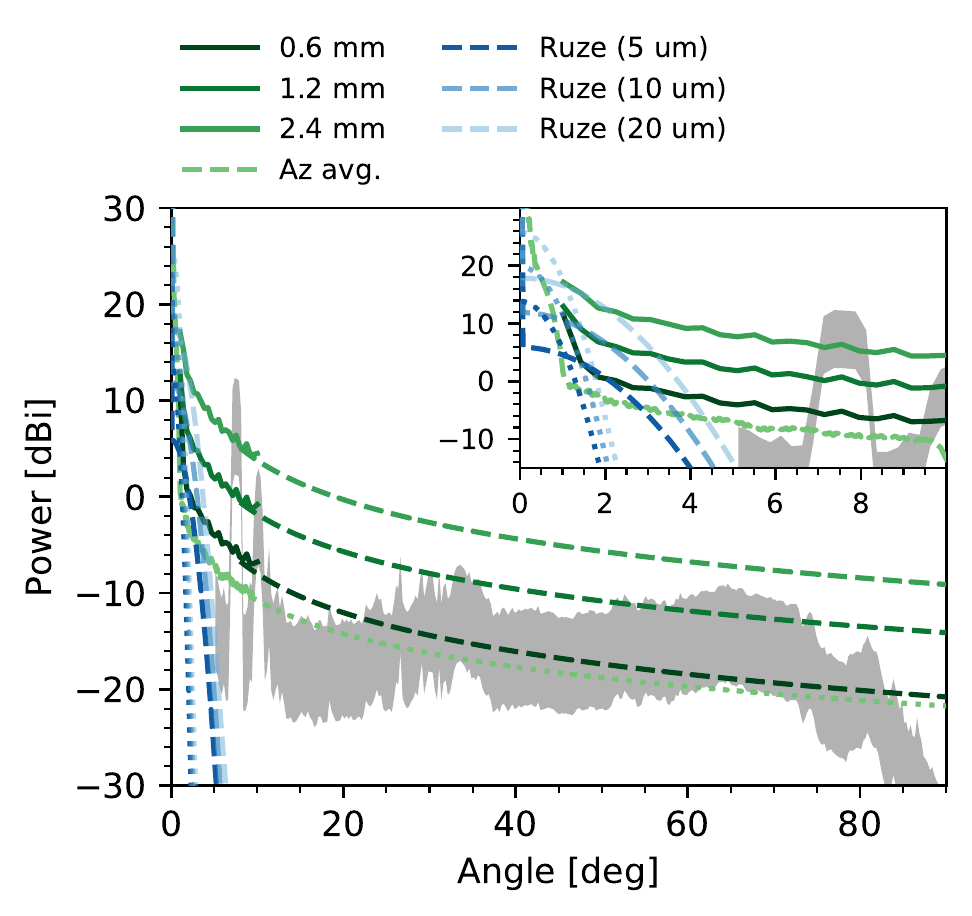}
    \caption{Predicted \SI[number-unit-product=\text{-}]{150}{\giga\hertz} beam profiles derived from panel gap simulations (green curves) and the spillover simulation (gray region) in Section \ref{sec:spillover}. The solid green lines, which have been truncated at \ang{1} for clarity, correspond to the average beam profile contained within the smaller region bounded by the dotted white lines in Fig.~\ref{fig:panel_gap_beam_map} (the arm of the diffraction pattern centered around elevation angle \ang{0}). The different shades of green correspond to profiles obtained when the panel gap separation is set to 0.6-, 1.2-, and \SI[number-unit-product=\text{-}]{2.4}{\milli\metre} while the light-green colored dashed line (labeled as "Az avg") corresponds to the full azimuthally-averaged beam profile for \SI[number-unit-product=\text{-}]{1.2}{\milli\meter} panel gap separation. All green curves have been extrapolated with a best-fit power law outside of \ang{10}. The gray region corresponds to the azimuthally averaged beam profiles obtained from the spillover analysis of the conical receiver baffle discussed in Section \ref{sec:spillover}. The upper and lower limits of the shaded region represent 3 and \SI{0.3}{\percent} spillover past the secondary, respectively. The gray shaded region has been truncated at \ang{7} because of sporadic (noisy) ray tracing results in the 1-\ang{7} region. The forward gain of the \SI[number-unit-product=\text{-}]{150}{\giga\hertz} beam is roughly \SI{78.0}{\deci\beli}. For reference, the blue dashed lines correspond to the predicted Ruze envelopes for a \SI[number-unit-product=\text{-}]{2}{\centi\meter} correlation length and a surface RMS of 5, 10, and \SI{20}{\micro\meter} while the dotted blue lines are the same for a \SI[number-unit-product=\text{-}]{5}{\centi\meter} correlation length. The beam profiles at \SI{90}{\giga\hertz} (omitted here for clarity) are quite similar in shape and amplitude. The inset panel shows a zoom in on the 0-\ang{10} region.}
    \label{fig:beam_profile}
\end{figure}



This model does not factor in scattering from the aperture of the elevation housing, which will likely impact far-sidelobes. However, the purpose of this study is to gauge the expected amplitude and angular shape of panel-gap diffraction lobes. A more advanced model that incorporates the shutter aperture is left for future work.

Figure \ref{fig:panel_gap_beam_map} shows the predicted far-field beam map at \SI{150}{\giga\hertz}. Diffraction from panel gaps produces an extended cross-shaped pattern with an approximate amplitude of \SI{20}{\deci\beli} at \ang{1} angle away from the beam center which decreases gradually (see Figure \ref{fig:beam_profile}). Although the feature is quite pronounced in the beam map, the fractional beam solid angle outside a 1-degree radius from the main beam centroid is less than \SI{0.1}{\percent} of the total predicted at both 90 and \SI{150}{\giga\hertz}. The amplitude of this diffraction feature is of course frequency dependent, the average beam profile amplitude in the 2-\ang{10} region is 1.0 and \SI{2.0}{\deci\beli} at 90 and \SI{150}{\giga\hertz}, respectively. In contrast, the azimuthally averaged beam profiles in that same angular region have a mean of -9.3 and \SI{-7.1}{\deci\beli} for 90 and \SI{150}{\giga\hertz}, respectively; significantly smaller than the cross-shaped diffraction features. The forward gain of a 90 and \SI{150}{\giga\hertz} pixel are predicted to be roughly 73.6 and \SI{78.0}{\deci\beli} at 90 and \SI{150}{\giga\hertz}, respectively. The expected beam profile amplitude from Ruze scattering \cite{Ruze1966}, assuming \SI[number-unit-product=\text{-}]{2}{\centi\meter} correlation length and 5-\SI{25}{\micro\meter} RMS error, is smaller than the azimuthally-averaged beam profiles predicted by these PO simulations at angles greater than \ang{5}. 

It is interesting to study the scaling of this panel-gap diffraction pattern with both frequency and panel separation. For this purpose, we have run these simulations using three values for panel gap separation corresponding to 0.6, 1.2 (nominal), and \SI{2.4}{\milli\metre}. Those simulations suggest that the amplitude of the beam sidelobes caused by panel gap diffraction depends on the physical separation between segments, as expected. For example, at 150 GHz, the predicted sidelobe amplitude at \ang{5} is -3.4, 2.6, \SI{8.4}{\deci\beli} for 0.6, 1.2, and \SI{2.4}{\milli\metre}, respectively. Figure \ref{fig:beam_profile} compares the panel gap diffraction beam profiles to that of the Ruze envelope for a single mirror with 2- and \SI[number-unit-product=\text{-}]{5}{\centi\meter} correlation length and a surface RMS of 5, 10, and \SI{20}{\micro\meter}. It is clear that the Ruze envelope model is unable to produce extended high amplitude sidelobes that are comparable to the panel gap diffraction pattern using correlation lengths and RMS errors that are consistent with existing panel surface RMS measurements (see Section \ref{sec:design}). We note that panel misalignment errors will most likely be driven by correlation lengths that are comparable or a factor of few smaller than the panel dimensions and that these deformations will be dominated by thermal and gravitational effects \citep{Parshley2018b}.

Figure \ref{fig:beam_profile} also compares the panel gap diffraction results to the predictions of the ray tracing spillover analysis presented in Section \ref{sec:spillover}. It is important to note that neither the spillover sidelobes nor the sidelobes from panel gap diffraction are expected to be azimuthally symmetric. Therefore, comparison of beam profiles as presented in Figure \ref{fig:beam_profile} only captures the relative power of these two effects after applying an azimuthal average. The sidelobe response will be dominated by different optical effects depending on both the radial and azimuthal sky-location relative to the beam centroid.

%% file: Conclusion/conclusion.tex
\section{Conclusions}

We have presented a suite of geometrical and physical optics simulations that shed light on the SO LAT optical design and some of its associated systematics. The simulation framework has informed all aspects of the optical design and we expect that the tools have been developed will also play a crucial role in assessing initial system performance when the telescope becomes operational. 

Through non-sequential ray tracing we have confirmed that absorbing material should be prioritized near the front (sky side) of the optics tubes. Non-sequential ray tracing has also shown that a simple conical baffle mounted in front of the telescope receiver helps direct near field sidelobes to the sky as quickly as possible and therefore minimizes loading from the \SI{300}{\kelvin} elevation housing. This analysis showed that little was gained from a more complex (and costly) parabolic baffle design. This additional baffle directs rays that would otherwise be distributed asymmetrically over wide angles to a more azimuthally symmetric pattern at roughly 7-\ang{8} from the main beam. 

We show results from physical optics simulations that suggest relatively compact and symmetric far field beams will be achieved. As expected, the physical optics simulations correlate significantly with Strehl ratio calculations from ray tracing. The physical optics calculations also show how the cold optics spillover is expected to vary across the focal plane under ideal conditions. Finally, we present physical optics simulations that address diffraction effects from panel gaps. These simulations show how the amplitude of the panel gap sidelobes is expected to change with frequency and panel gap separation.

The shape of the far field beam maps of the Simons Observatory Small and Large Aperture telescopes impact all astrophysical and cosmological analysis made by the experiment. In general, the non-symmetric part of a beam response can couple parity even and parity odd polarization modes on the sky and therefore complicate the interpretation of Stokes $Q$ and $U$ sky maps. For the SO LAT, beam asymmetries can impact the shape reconstruction of both the CMB lensing potential and extended objects such as galaxy clusters. Simulation work that sheds light on the impact of beam non-idealities is currently in progress and will be discussed in future publications. The extensive simulation capabilities that we have developed allow us to model the as-built instrument and compare the expected and realized signal response, which will continue to aid in optimizing the observatory performance over time.

Throughout the SO LAT design process we have tried to optimize for both beam symmetry and forward gain. We have also explored how several factors contribute to beam spillover and affect the mapping speed of the instrument by coupling the detectors to unwanted radiation sources, such as the inside of the telescope structure, a distant mountain, or the moon. Measurements of spillover and scattering from instrument optics (lenses, filters, and baffles) are now getting underway; this will enable initial model comparisons. Once the instrument is deployed on the telescope, one of the first goals will be to test these predictions under a wide range of conditions through observations of point sources such as planets. These results will later be used for calibration and to aid with interpretation of the first cosmological measurements with the SO LAT.

